\documentclass[]{aa} % for a referee version

\usepackage{graphicx}
\usepackage{txfonts}
\begin{document}

\title{Galactic cold cores
II. Herschel study of the extended dust emission around the first Planck detections
\thanks{
{\it Planck} \emph{(http://www.esa.int/Planck)} is a project of the European Space
Agency -- ESA -- with instruments provided by two scientific consortia funded
by ESA member states (in particular the lead countries: France and Italy) with
contributions from NASA (USA), and telescope reflectors provided in a
collaboration between ESA and a scientific Consortium led and funded by
Denmark.
}
\thanks{{\it Herschel} is an ESA space observatory with science instruments provided
by European-led Principal Investigator consortia and with important
participation from NASA.}
}

\author{M.     Juvela\inst{1},
        I.     Ristorcelli\inst{2},
        V.-M.  Pelkonen\inst{3},
        D.J.   Marshall\inst{2},
        L.A.   Montier\inst{2},
        J.-P.  Bernard\inst{2},
        R.     Paladini\inst{3},
        T.     Lunttila\inst{1},
        %%%%%%%%%%%%%%%%%%%%%%%%%%%%%%%%%%%%%%%
        A.     Abergel\inst{4},
        Ph.    Andr\'e\inst{5},
        C.     Dickinson\inst{6},
        X.     Dupac\inst{7},
        J.     Malinen\inst{1},
        P.     Martin\inst{8},
        P.     McGehee\inst{3},
        L.     Pagani\inst{9},        
        N.     Ysard\inst{1},
        A.     Zavagno\inst{10}
        }

\institute{
Department of Physics, P.O.Box 64, FI-00014, University of Helsinki,
Finland, {\em mika.juvela@helsinki.fi}
\and
CESR, Observatoire Midi-Pyr\'en\'ees (CNRS-UPS), Universit\'e de
Toulouse, BP 44346, 31028 Toulouse Cedex 04, France 
\and
IPAC, Caltech, Pasadena, USA
\and
IAS, Universit\'e Paris-Sud, 91405 Orsay cedex, France
\and     
Laboratoire AIM, CEA/DSM­CNRS­Universit\'e Paris Diderot,
IRFU/Service d'Astrophysique, CEA Saclay, 91191 Gif-sur-Yvette, France
\and
Jodrell Bank Centre for Astrophysics, University of Manchester,
Oxford Road, Manchester, M13 9PL, U.K.
\and
European Space Agency, European Space Astronomy Centre, Villanueva de
la Ca\~nada, Madrid, Spain 
\and
Canadian Institute for Theoretical Astrophysics (CITA),
60 St. George St., Toronto ON M5S 3H8, Canada
\and
LERMA \& UMR 8112 du CNRS, 
Observatoire de Paris, 61 Av. de l'Observatoire, 75014 Paris, France
\and
Laboratoire d'Astrophysique de Marseille,
38 rue F. Joliot-Curie, 13388 Marseille Cedex 13, France
}

\authorrunning{M. Juvela et al.}

\date{Received September 15, 1996; accepted March 16, 1997}

\abstract
   {
   Within the project $Galactic$ $Cold$ $Cores$ we are carrying out
   Herschel photometric observations of cold interstellar clouds 
   detected with the Planck satellite. The three fields observed as
   part of the Herschel science demonstration phase (SDP) provided the
   first glimpse into the nature of these sources. The aim of the
   project is to derive the physical properties of the full cold core
   population revealed by Planck. 
   }
   {
   We examine the properties of the dust emission within the three
   fields observed during the SDP. We determine the dust
   sub-millimetre opacity, look for signs of spatial variations in
   the dust spectral index, and estimate how the apparent variations
   of the parameters could be affected by different sources of
   uncertainty.
   }
   {
   We use the Herschel observations where the zero point of the
   surface brightness scale is set with the help of the
   Planck satellite data. We derive the colour temperature and
   column density maps of the regions and determine the dust
   opacity by a comparison with extinction measurements. By
   simultaneously fitting the colour temperature and the dust spectral
   index values we look for spatial variations in the apparent dust
   properties. With a simple radiative transfer model we estimate to
   what extent these can be explained by line-of-sight temperature
   variations, without changes in the dust grain properties.
   }
   {
   The analysis of the dust emission reveals cold and dense clouds
   that coincide with the Planck sources and confirm those detections.
   The derived dust opacity varies in the range
   $\kappa(250\,\mu{\rm m}) \sim$ 0.05--0.2\,cm$^2$\,g$^{-1}$, higher
   values being observed preferentially in regions of high column
   density. The average dust spectral index $\beta$ is $\sim$1.9--2.2.
   There are indications that $\beta$ increases towards the coldest
   regions. The spectral index decreases strongly near internal
   heating sources but, according to radiative transfer models, this
   can be explained by the line-of-sight temperature variations
   without a change in the dust properties.
   }
   {}


   \maketitle
%
%________________________________________________________________

\begin{figure}
\centering
\includegraphics[width=8cm]{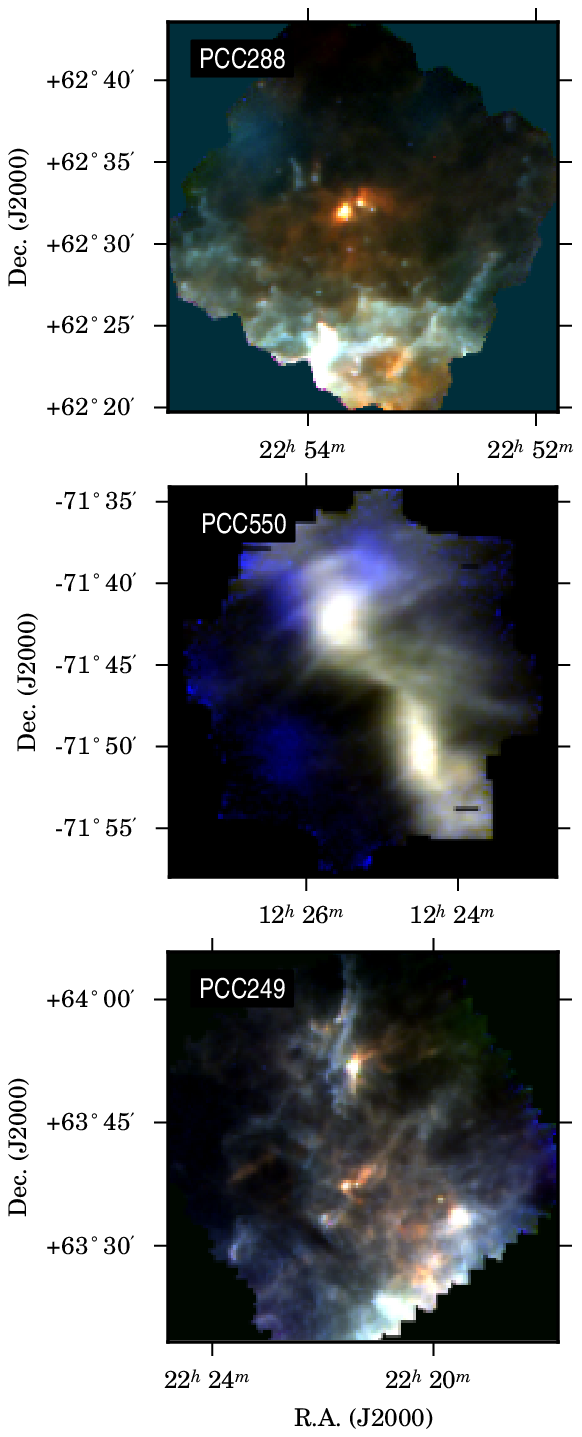}
\caption{
Three-colour figures of the three SDP fields. The red, green, and blue
colours correspond to the 250\,$\mu$m, 160\,$\mu$m, and 100\,$\mu$m
intensities (PCC288 and PCC249) or the 350\,$\mu$m, 250\,$\mu$m, and
160\,$\mu$m intensities (PCC550). Only the area covered by both the PACS and
SPIRE instruments is shown. See Appendix~\ref{sect:frequency_maps} for
images of the individual bands. 
}
\label{fig:rgb}%
\end{figure}

\section{Introduction}

A major question in star formation studies is how the main features of
the process depend on the initial conditions in cold molecular clouds.
The masses of the formed stars, the star formation efficiency, the
differences between clustered and isolated star formation, and the
timescales are all closely linked to the properties of dense cloud
cores. In order to understand these dependencies, we need to study a
large sample of molecular clouds and the pre-stellar cores within
them.  The molecular lines are a central tool in the study of
molecular clouds. However, in the cloud cores many molecules have
frozen onto dust grains making it difficult to carry out quantitative
analysis with molecular line data alone (e.g., Kramer et al.
\cite{Kramer1999}). One has to resort to observations of the dust
component of the clouds, either using the extinction or the dust
emission as a tracer of the mass distribution.

The recently launched Herschel and Planck satellites represent a
significant step forward in sub-millimeter studies by eliminating the
atmospheric effects from observations. The satellites enable sensitive
measurements of the dust emission in Galactic interstellar clouds,
Planck over the whole sky and Herschel in smaller regions but with
higher angular resolution. The satellites provide the complementarity
needed for a study of a full cross section of the Galactic cold core
population. Because of its high sensitivity and the coverage of the
sub-millimetre wavelengths dominated by cold dust emission, the Planck
all-sky maps provide an unbiased census of cold dust clouds and, in
particular, of compact pre-stellar cloud cores. The survey of cold
compact dust clouds is one of the goals in the Galactic science
programme of Planck. In the Herschel Open Time Key Programme {\em
Galactic Cold Cores}, we use the Herschel PACS and SPIRE
instruments to map about one hundred target fields that are selected on the
basis of the Planck survey. This study was begun during the Herschel
Science Demonstration Phase (SDP) when three fields were observed.

\begin{table}
\caption{The observed fields}
\label{table:obs}  
\centering         
\begin{tabular}{c c c c}
\hline\hline            
Target     &  RA (J2000)       & DEC (J2000)  &   Map size (PACS/SPIRE)\\
\hline         
 PCC\,249  &    22 21 17.6   &   +63 42 25    &  50$\arcmin$/50$\arcmin$  \\
 PCC\,288  &    22 53 31.3   &   +62 31 44    &  18$\arcmin$/30$\arcmin$  \\
 PCC\,550  &    12 25 16.5   &   $-$71 46 03    &  18$\arcmin$/30$\arcmin$  \\
\hline
\end{tabular}
\end{table}

The main features of the compact sources found in the SDP fields were
reported in Juvela et al. (\cite{Juvela2010}). In this paper we
concentrate on the dust emission at large scales. We are interested in
the cloud properties, their temperature and column density
distributions. However, we also want to study the dust component
itself. These are not unrelated questions because the derived cloud
masses depend on the assumptions of the dust opacity and spectral
index. Conversely, the variations in dust properties should be related
to the physical conditions within the clouds.

The dust spectral index and its temperature dependence have received
much attention because they provide additional information on the
chemical composition, structure, and the size distribution of
interstellar dust grains (e.g., Ossenkopf \& Henning
\cite{ossenkopf1994}; Krugel \& Siebenmorgen \cite{Krugel1994};
Mennella et al. \cite{Mennella1998}; Boudet et al. \cite{Boudet2005};
M\'eny et al. \cite{Meny2007}; Compi\`egne et al.
\cite{Compiegne2011}). The observations consistently show an inverse
$T-\beta$ relation (e.g., Dupac et al.~\cite{Dupac2003}, Hill et
al.~\cite{Hill2006}, Veneziani et al.~\cite{Veneziani2010} and the
initial Herschel results including Anderson et al. \cite{Anderson2010}
and Rodon et al. \cite{Rodon2010}).
However, the reliability of this relation is notoriously difficult to
estimate because any noise present in the measurements produces a
similar anticorrelation (Schwartz et al. \cite{Schwartz1982}, Dupac et
al. \cite{Dupac2003}, Shetty et al.~\cite{Shetty2009a}).  Shetty et
al. (\cite{Shetty2009b}) studied the importance of observational noise
and temperature variations on the reliability of the derived dust
properties. In the presence of line-of-sight temperature variations,
the observed dust spectral index $\beta$ will underestimate the real
spectral index of the dust grains. The estimated colour temperatures
are not only biased towards the warm regions but, for example in the
two component models of Shetty et al. (\cite{Shetty2009b}), could be
even higher than any of the real dust temperatures. Similar results
were obtained by Malinen et al. (\cite{Malinen2011}) who studied
synthetic observations produced by combining MHD modelling and
radiative transfer calculations. They concluded that the presence of
protostellar radiation sources can further increase the bias in the
observed $\beta$ values and can produce an apparent inverse
$\beta$--$T$ relation that is difficult to separate from any intrinsic
relation the dust grains may have. Malinen et al. (\cite{Malinen2011})
also noted that when the spectral index varies over the wavelength
range included in the spectral energy density (SED) fit, the derived
spectral index can rise even above the real $\beta(\lambda)$ of the
dust anywhere in that interval. Because an error in the spectral index
is always associated with a compensating error in the temperature,
this affects the mass estimates and could even bias the derived
density profiles of compact objects.

Like the spectral index, also the absolute value of the dust
absorption cross section, $\kappa$, is expected to be variable. This
could be caused by a change in the abundance of different dust
populations, a change in the grain size distribution (see, e.g.,
Steinacker et al. \cite{Steinacker2010}), or a change in the
composition of individual grains. For the sub-millimetre observations,
the effects should again be strongest in dense and cold regions where
the grains acquire ice mantles and may more easily coagulate to form
larger grains with $\kappa$ values up to 3--4 times higher (e.g.,
Ossenkopf \& Henning \cite{ossenkopf1994}, Krugel \& Siebenmorgen
\cite{Krugel1994}, Stepnik et al. \cite{Stepnik2003}). An error in
$\kappa$ automatically translates to a similar fractional error in the
mass estimates. The absolute value of $\kappa$ is not easy to measure
because it requires an independent column density estimate that is
equally or more reliable over the $A_{\rm V}$ range for which
sub-millimetre observations exist.

In this paper we study these questions with the help of photometric
Herschel observations. We start by deriving the colour temperature
(Sect.~\ref{sect:TC}) and column density (Sect~\ref{sect:N}) maps of
the three SDP fields. This first analysis is done with a constant
value of the spectral index $\beta$. In Sect.~\ref{sect:beta} we
proceed to study the evidence for variations in the dust spectral
index. With simultaneous fits of colour temperature and $\beta$, we
calculate maps of the apparent spectral index paying special attention
to the error sources that could affect the obtained values or the
morphology of the $\beta$ maps.  By comparing the results with the
near-infrared reddening of background stars, we calculate maps of dust
sub-millimetre opacity (Sect.~\ref{sect:emissivity}).  To
complement this analysis, we present in Sect.~\ref{sect:model} a
simple radiative transfer model for the field PCC288 and use the
results to estimate the difference between the apparent $\beta$ and
the real spectral index of the dust grains. After a discussion in
Sect.~\ref{sect:discussion} we present our final conclusions in
Sect.~\ref{sect:conclusions}.

\section{Observations}  \label{sect:obs}

The Planck satellite is performing all sky surveys at nine
wavelengths between 350\,$\mu$m and 1\,cm (Tauber et al.
\cite{Tauber2010}). With the detection methods described in Montier et
al. (\cite{Montier2010}), three target fields were selected from the
Planck First Light Survey (see Table~\ref{table:obs}) for photometric
observations with the Herschel PACS and SPIRE instruments (for details
about Herschel, see Pilbratt et al. \cite{Pilbratt2010}, Poglitsch et
al. \cite{Poglitsch2010}, and Griffin et al. \cite{Griffin2010}). The
Herschel observations were performed in November and December 2009.
The field PCC249 was observed in the parallel mode and the fields
PCC288 and PCC550 in the normal scan mapping, one instrument at a time
(see Juvela et al. \cite{Juvela2010}). The fields are shown in
Fig.~\ref{fig:rgb}.
The data were reduced with the {\it Herschel} Interactive Processing
Environment (HIPE) v.2. The PACS maps were created using the madmap
algorithm. The processing includes some high pass filtering. In order
to minimize the removal of large scale structure, the length of the
filter window was kept equal to the scan length. The SPIRE maps are
the product of direct projection onto the sky and averaging of the
time ordered data.

\begin{table}
\caption{The uncertainties of the map zero levels}
\label{table:offsets}  
\centering         
\begin{tabular}{c c c c}
\hline\hline            
Target     &    Wavelength    &   Zero level uncertainty   \\
           &    ($\mu$m)      &   MJy\,sr$^{-1}$           \\
\hline         
 PCC\,288  &     100          &  14.3      \\
           &     160          &  10.3      \\
           &     250          &   3.1      \\
           &     350          &   1.2      \\
           &     500          &   0.4      \\
 PCC\,550  &     100          &   0.8$^1$  \\
           &     160          &   1.4$^1$  \\
           &     250          &   0.43     \\
           &     350          &   0.23     \\
           &     500          &   0.08     \\
 PCC\,249  &     100          &   4.4      \\
           &     160          &   2.6      \\
           &     250          &   2.9      \\
           &     350          &   0.7      \\
           &     500          &   0.21     \\
\hline         
\multicolumn{3}{l}{$^1$Zero level calculated using map averages.}
\end{tabular}  
\end{table}

In order to study dust properties over the whole fields we need to
establish a zero point for the intensity scales. We use the Planck
maps (data release DX4, available within the Planck core teams) at
350\,$\mu$m and 550\,$\mu$m and the Improved Reprocessing of the IRAS
Survey (IRIS, Miville-Deschenes \& Lagache \cite{MAMD2005}) maps at
100\,$\mu$m. The highest frequency Planck channels have been
calibrated with FIRAS (Piat et al. 2002) and the IRIS data have been
calibrated according to DIRBE. 
When necessary, we interpolate these reference data for the Herschel
bands using a modified black body curve $B_{\nu} \nu^2$ fitted to the
nearest two wavelength points. 

The 100\,$\mu$m band contains a contribution from very small grains
(VSG). The effect is approximately similar in the IRIS and PACS
100\,$\mu$m bands and does not affect the calibration comparison at
that wavelength. However, the presence of VSG at 100\,$\mu$m could
bias the values interpolated to 160\,$\mu$m and 250\,$\mu$m.
Therefore, before the interpolation, we subtract from the IRIS
100\,$\mu$m data the VSG contribution estimated with the D\'esert et
al. (\cite{Desert1990}) dust model. The radiation field is scaled so
that the model reproduces the observed temperature of the big grains
(BG) that dominate the emission wavelengths longer than 100\,$\mu$m.
The predicted ratio of VSG and BG emission at 100\,$\mu$m is used to
correct the IRIS data. Because the ratio depends on the estimated BG
temperature, the correction has to be done iteratively.
Finally, we do a least squares fit to Herschel surface brightness
(convolved to a resolution of 4.3$\arcmin$) vs. Planck/IRIS and use
the offset to set the zero level of the Herschel maps while keeping
the original gain calibration.  For the PCC550 PACS observations the
fits remained uncertain because of the small amount of surface brightness
variation. In this case the zero point was estimated using the
difference in the average surface brightness rather than the
extrapolation to zero Planck/IRIS surface brightness.

In the comparison between Planck and IRIS, the data should be colour
corrected because this could affect the quality of the least squares
fit. The colour corrections were calculated for modified black body
spectra $B_{\nu}(T) \nu^2$ that were fitted to the observations after
a preliminary estimation of the zero points. All observations,
including Planck and IRIS data, were colour corrected before repeating
the least squares fits. The corrections can be applied consistently
only if the offsets of the Herschel and Planck/IRIS maps are already
correct and therefore also this correction needs to be done
iteratively. However, the colour corrections are less than 10\%, and a
sufficient accuracy is reached already on the second iteration. The
subsequent analysis was carried out using these colour corrected
observations.

The statistical uncertainties of the offsets were estimated with the
bootstrap method and are listed in Table~\ref{table:offsets}. The
uncertainties are of the order of 1\,MJy\,sr$^{-1}$ but smaller at the
longest wavelengths with the lowest surface brightness. The error
estimates are larger for the PACS channels apart from the field PCC550
where, as explained above, the quoted offsets are calculated using
average surface brightness values.

\section{Observational results}  \label{sect:results}

In this section we present the colour temperature, column density, and
spectral index maps. The relation between these and the intrinsic dust
properties is discussed in the later sections.

\subsection{Colour temperature maps} \label{sect:TC}

The colour temperature maps of the large grain emission were estimated
using the 100\,$\mu$m, 160\,$\mu$m, 250\,$\mu$m, 350\,$\mu$m, and
500\,$\mu$m observations.  The colour temperatures were first
calculated with a constant value of the spectral index, $\beta=2.0$.
The validity of this approximation is investigated in
Sect.~\ref{sect:beta}. The use of a constant $\beta$ can be justified
even when the assumption is not strictly true because it enables more
robust estimation of {\em one} colour temperature that characterizes
the fields. As discussed in Sect.~\ref{sect:model}, the colour
temperature is often an imprecise tracer of the physical dust
temperatures.

\begin{figure*}
\centering
\includegraphics[width=!]{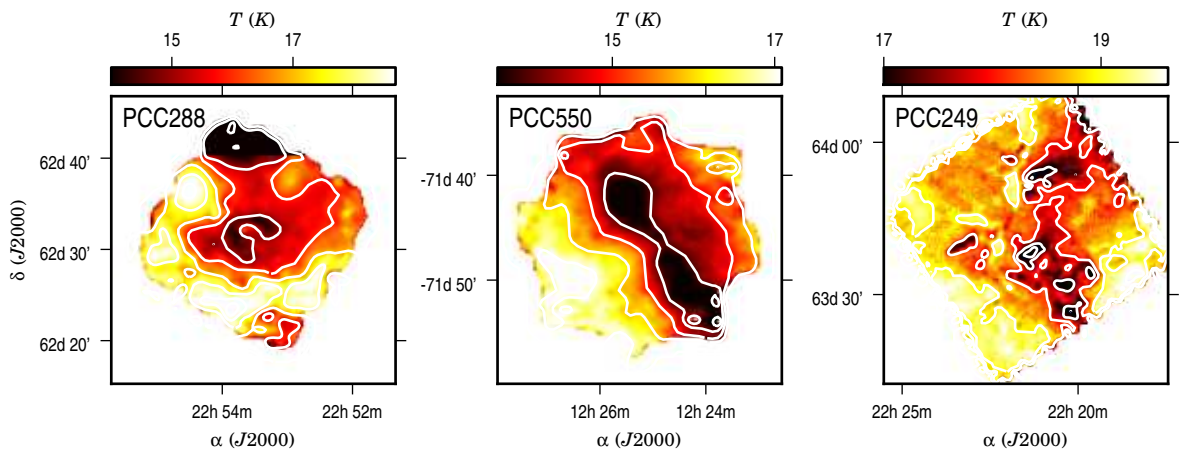}
\caption{
Dust colour temperature maps for the three SDP fields. The
calculations assume a fixed value of the spectral index, $\beta=2.0$.
The map resolution is $\sim$37$\arcsec$ and the contours are drawn 
one Kelvin apart.
}
\label{fig:T_beta2}%
\end{figure*}

All data were smoothed to the resolution of the SPIRE 500\,$\mu$m map
(37$\arcsec$) using a Gaussian kernel. As discussed in
Sect.~\ref{sect:obs}, the determination of the zero points of the
surface brightness scales and the correction for the VSG contribution
were done iteratively together with the calculation of the colour
temperatures. In the fit the relative weight of the data are set
assuming 10\% uncertainty for the SPIRE and 20\% for the PACS
measurements. The temperature maps are shown in
Fig.~\ref{fig:T_beta2}.

\subsection{Column density maps} \label{sect:N}

\begin{figure*}
\centering
\includegraphics[width=!]{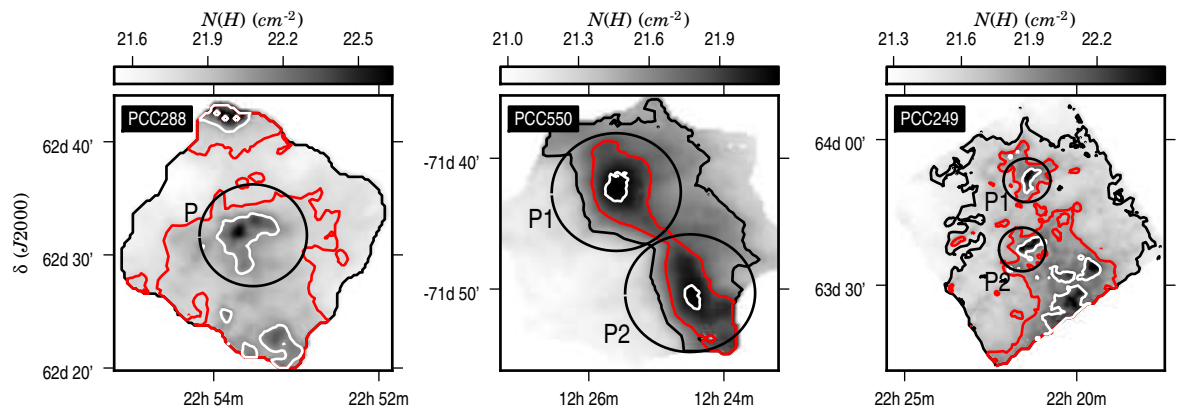}
\caption{
Maps of the hydrogen column density $N({\rm H})$ derived with the
colour temperature maps of Fig.~\ref{fig:T_beta2}. The maps have a
resolution of $\sim$37$\arcsec$ and the contours are drawn at levels
where $log_{\rm 10}N(H)$ equals 21.5, 21.8, and 22.1. The circles show
the 9 arcmin apertures used in Juvela et al. (\cite{Juvela2010}).
}
\label{fig:colden}
\end{figure*}

The column densities were calculated using the modified blackbody fits
that resulted in the colour temperature maps of
Fig.~\ref{fig:T_beta2}. Like the spectral index, the dust absorption
cross section may vary from region to region but the analysis was
performed using a constant value of
$\kappa_{850}$=0.02\,cm$^{2}$\,g$^{-1}$ (see Juvela et
al.~\cite{Juvela2010}). With a $\nu^2$ dependence this corresponds to
$\kappa_{250}$=0.23\,cm$^{2}$\,g$^{-1}$. The column density maps are
presented in Fig.~\ref{fig:colden} where we also show 9 arcmin
diameter circles around the positions of the Planck detections. These
were used in Juvela et al.~(\cite{Juvela2010}) to estimate the masses
of the regions.

The column densities were transformed into mass with the distances
adopted by Juvela et al. (\cite{Juvela2010}), 800\,pc, 225\,pc, and
800\,pc for PCC288, PCC550, and PCC249, respectively. In the same
order, the total masses of the areas displayed in
Fig.~\ref{fig:colden} are $\sim$890\,$M_{\sun}$, $\sim$31\,$M_{\sun}$,
and $\sim$3800\,$M_{\sun}$. The masses above the given column density
contours, and inside the 9 arcmin apertures, are listed in
Table~\ref{table:masses}. The statistical errors are small and
the mass uncertainty is dominated by the uncertainties of the
calibration ($\sim$10\%), the dust opacity (up to a factor of two),
and the cloud distance (typically $\sim$30\% corresponding to
a $\sim$50\% uncertainty in the mass.

\subsection{Dust spectral index} \label{sect:beta}

\begin{figure}
\centering
\includegraphics[width=8cm]{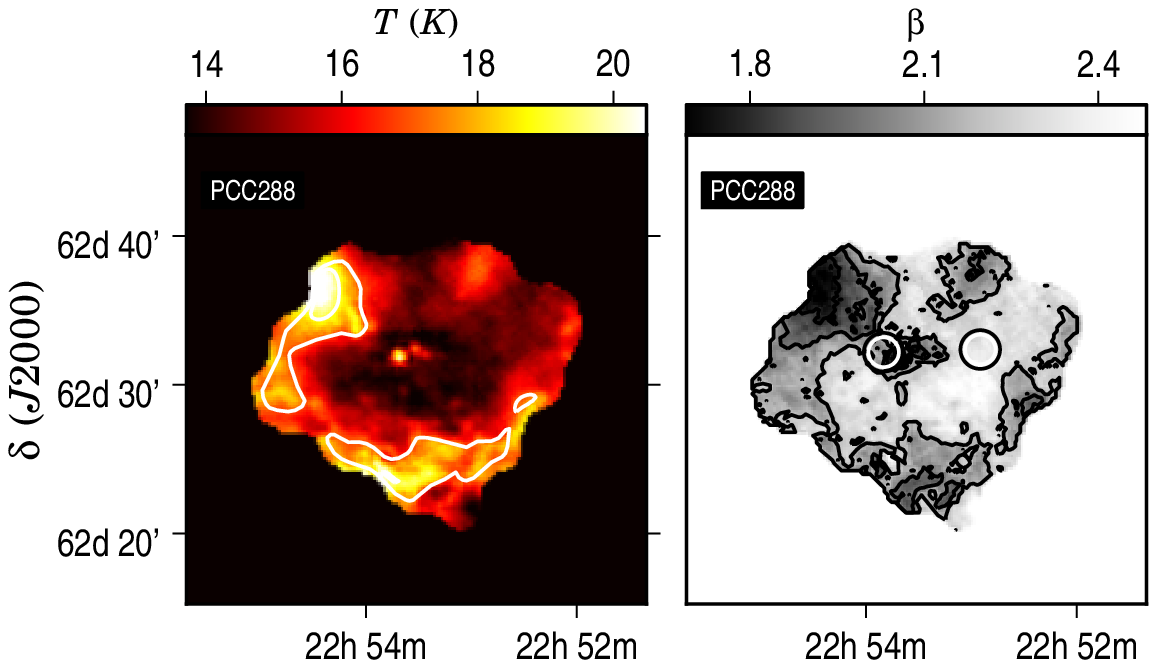}
\includegraphics[width=8cm]{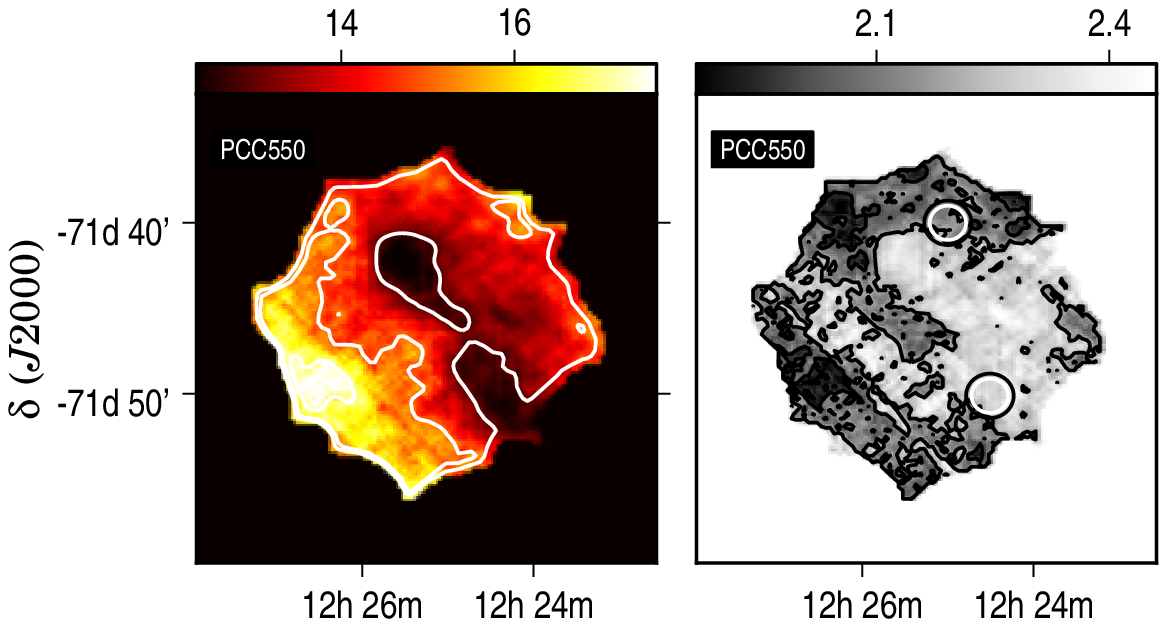}
\includegraphics[width=8cm]{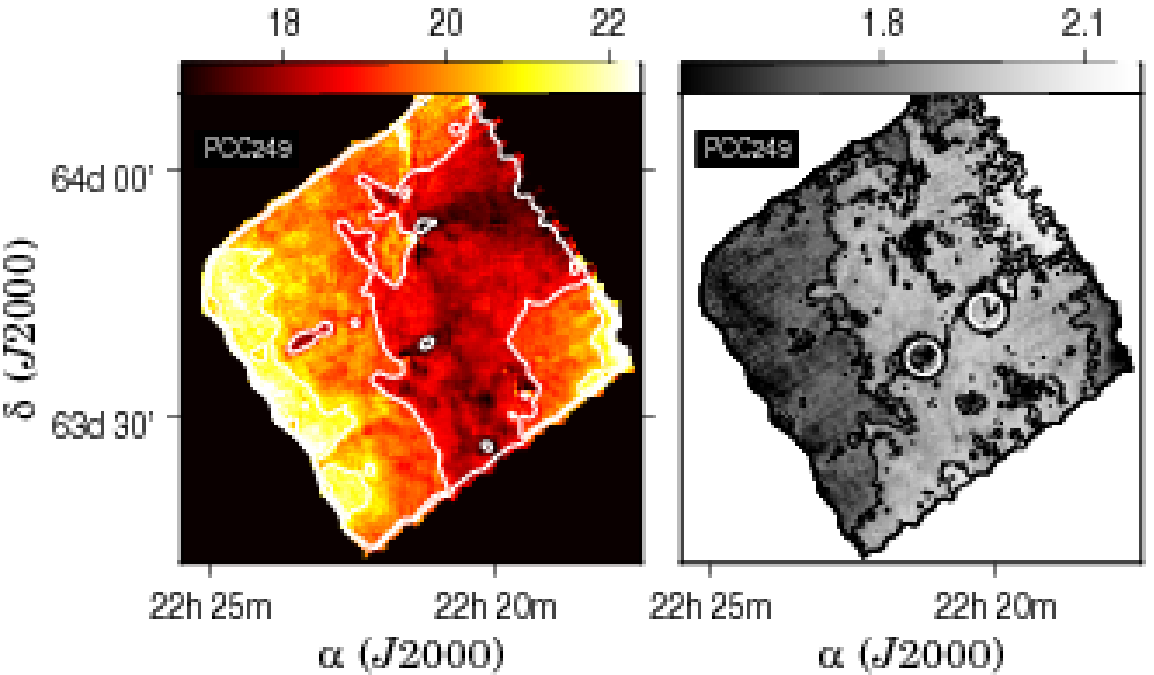}
\caption{
The colour temperature and dust spectral index maps. The values have
been determined by a fit to the observations between 100\,$\mu$m and
500\,$\mu$m. The circles indicate the positions for which the spectra
are shown in Fig.~\ref{fig:SEDs}.
}
\label{fig:T_beta}
\end{figure}

\begin{figure}
\centering
\includegraphics[width=8cm]{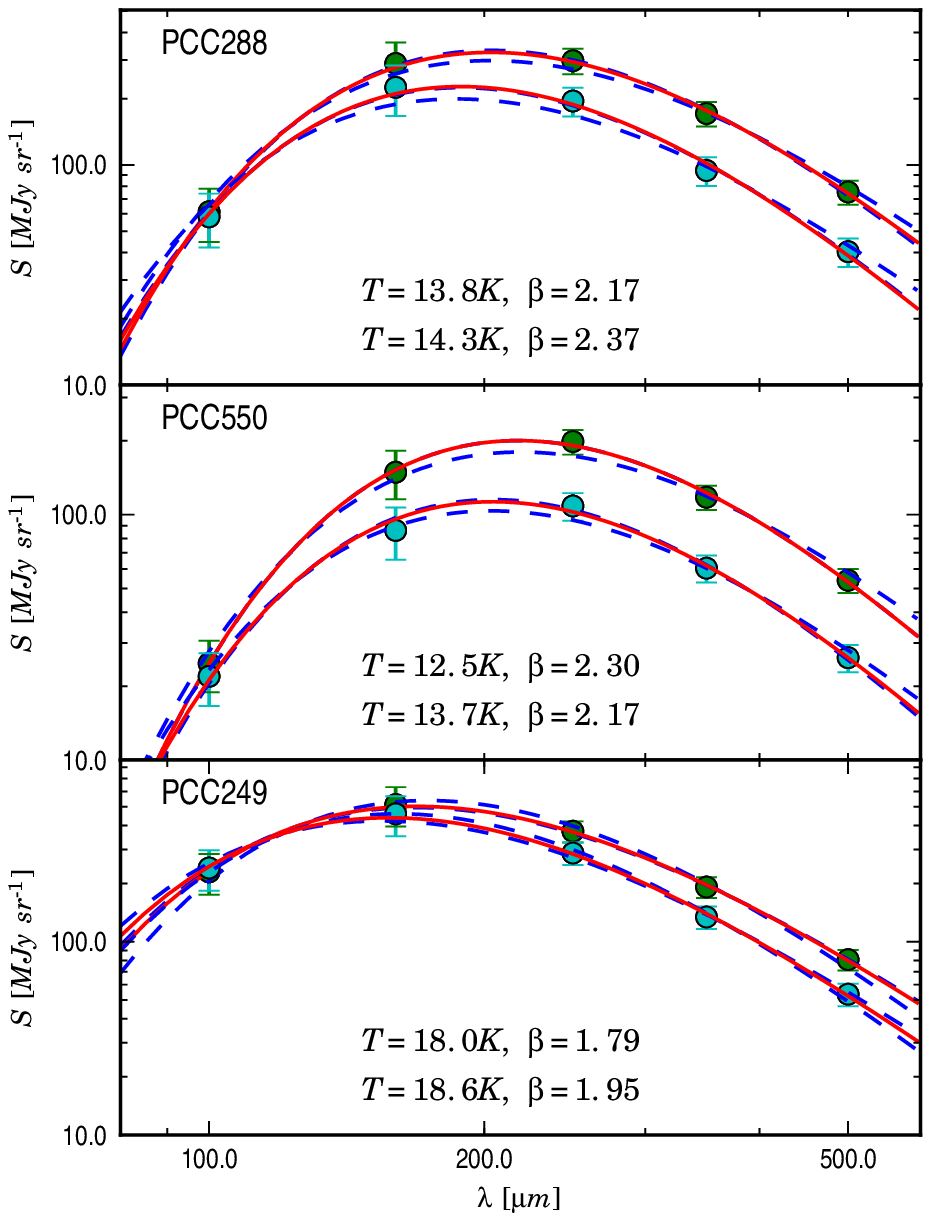}
\caption{
Spectral energy distributions for selected positions. The measurements
correspond to a single pixel at the centre of the circles shown in
Fig.~\ref{fig:T_beta} (after convolution to the resolution of the
500\,$\mu$m observations). The temperature and the spectral index are
quoted in the frames, the first values corresponding to the upper SED
curve. The dashed lines are the fits with a constant spectral index
value of $\beta=1.7$ or $\beta=2.3$.
}
\label{fig:SEDs}
\end{figure}

The 100\,$\mu$m--500\,$\mu$m data were also fitted with modified
blackbody curves $B_{\nu}(T) \nu^{\beta}$ keeping both the colour
temperature $T$ and the spectral index $\beta$ as free parameters.  
We masked out the borders of the maps ($\sim 3\arcmin$) and the
northern end of the field PCC288 where the column density map and the
map of dust opacity (see Sect.~\ref{sect:emissivity} and
Sect.~\ref{sect:disc_beta}) show a suspicious feature. The colour
temperature and spectral index maps are shown in Fig.~\ref{fig:T_beta}
and representative SEDs in Fig.~\ref{fig:SEDs}.
The median values of $\beta$ are 2.21, 2.21, and 1.86 for PCC288,
PCC550, and PCC249, respectively.
The same fits without the 100\,$\mu$m data are shown in 
Appendix~\ref{sect:T_beta_no_100}.

\begin{table}
\caption{%
The mass estimates of the observed fields or parts of the fields that
are defined by a column density limit or correspond to 9 arcmin
apertures (regions $P$, $P1$, and $P2$).
}
\label{table:masses} \centering 
\begin{tabular}{c c c}
\hline\hline            
Field      &    Region                 &  Mass ($M_{\sun}$)  \\
\hline         
 PCC\,288  &   full field              &  1108    \\
           &   $log_{\rm 10} N>21.8$   &   785    \\
           &   $log_{\rm 10} N>22.1$   &   242    \\
           &   $P$                     &   288    \\
 PCC\,550  &   full field              &    36.2  \\
           &   $log_{\rm 10} N>21.8$   &    13.9  \\
           &   $P1$                    &    12.2  \\
           &   $P2$                    &    11.3  \\           
 PCC\,249  &   full field              &  3802    \\
           &   $log_{\rm 10} N>21.8$   &  1763    \\
           &   $log_{\rm 10} N>22.1$   &    627   \\
           &   $P1$                    &    252   \\
           &   $P2$                    &    259   \\
\hline
\end{tabular}
\end{table}

\begin{figure*}
\centering
\includegraphics[width=5.5cm]{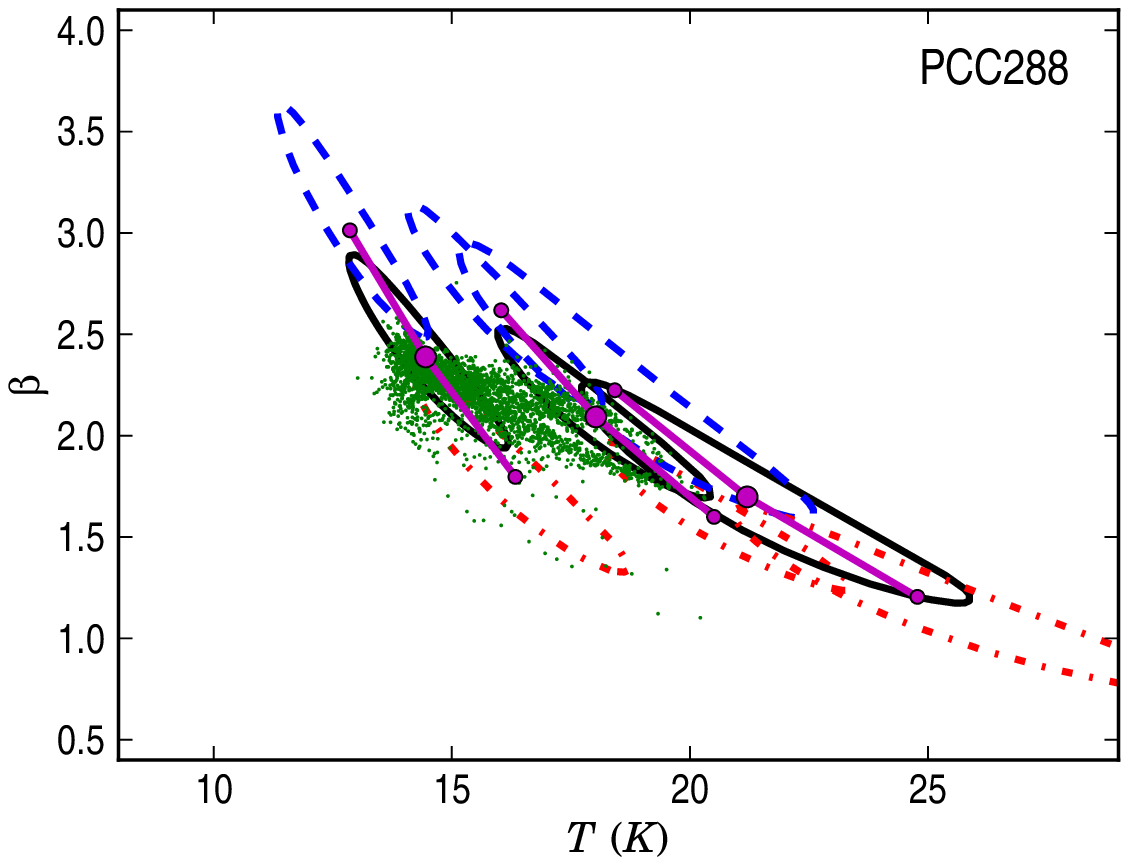}
\includegraphics[width=5.5cm]{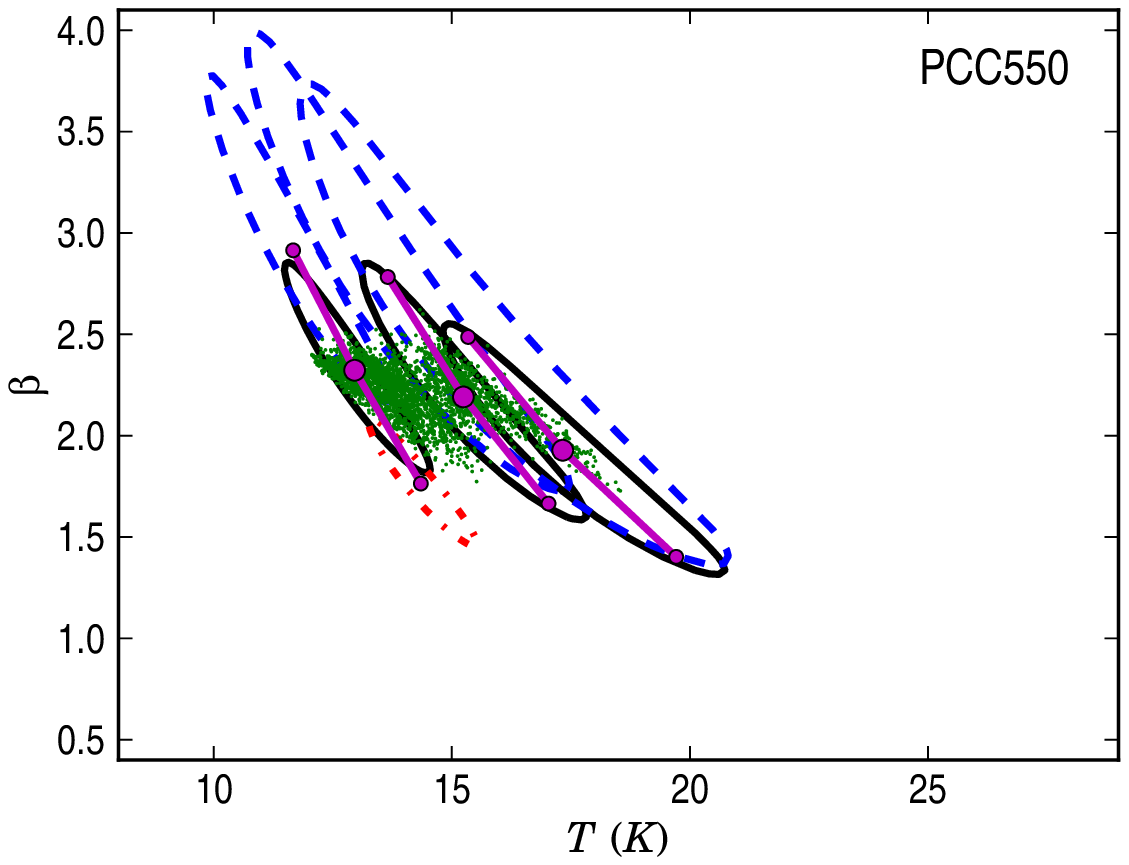}
\includegraphics[width=5.5cm]{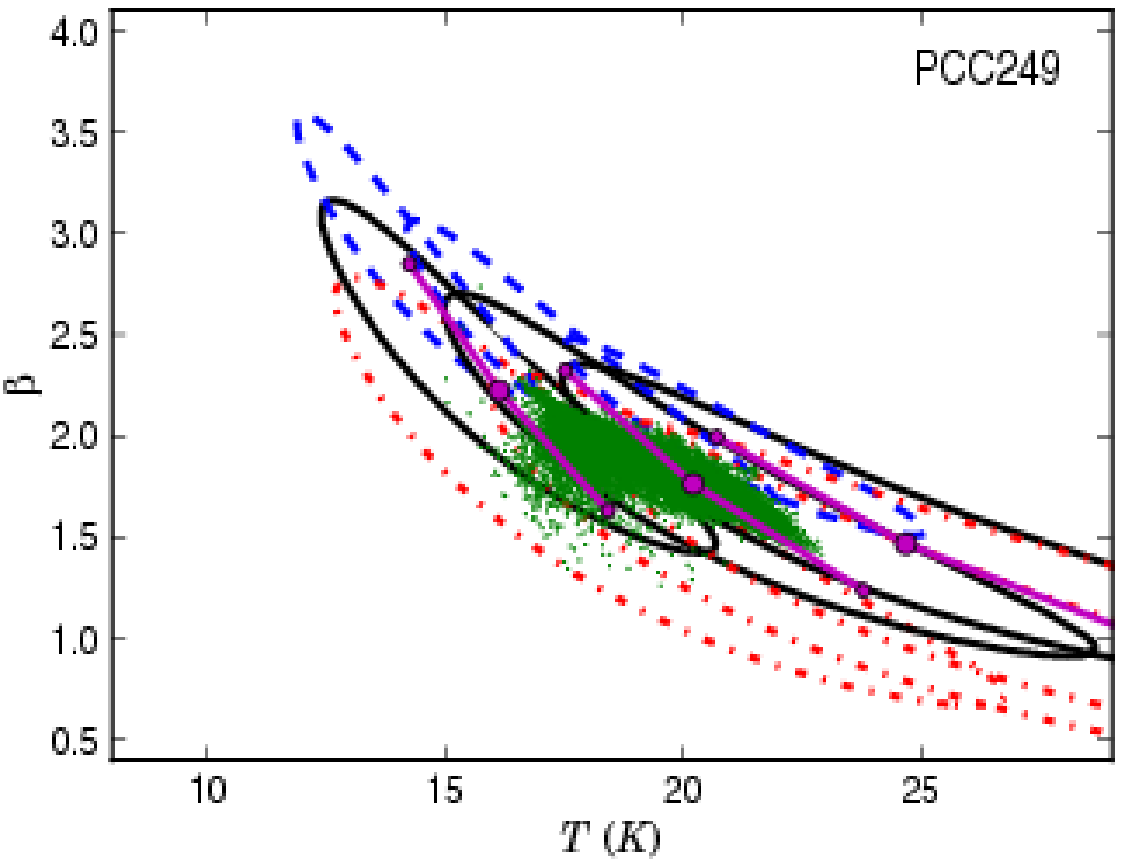}
\caption{
The observed temperature dependence of the spectral index in the three
fields. The values are shown for individual pixels (green dots). For
three selected pixels, the black contours show the limit at which the
$\chi^2$ value of the SED fit has doubled from its minimum value. For
the same pixels the smaller symbols (joined with solid lines) show the
extreme values allowed by the calibration uncertainty (see text). The
error regions for these extreme cases are drawn with dashed and
dash-dot lines.
}
\label{fig:T_beta_plot}
\end{figure*}

In Fig.~\ref{fig:T_beta_plot} we show the relations between the
derived spectral indices and the colour temperatures.  For three
randomly selected pixels, the black contours show where the $\chi^2$
value of the SED fit has doubled from its minimum value.  

To get the first idea of the possible effect of the calibration
uncertainty, we compared two cases where we add errors to the
250\,$\mu$m and the 500\,$\mu$m maps and recalculate the colour
temperature and spectral index maps. The wavelengths were chosen
because their relative intensities are important in determining the
spectral index. The map offsets were first changed by the values
listed in Table~\ref{table:offsets} and the surface brightness values
were scaled corresponding to a 10\% gain uncertainty. The changes at
the two wavelengths were made in opposite directions in order to have
the maximum effect in the derived $\beta$ values. By applying the
$\sim$1$\sigma$ shifts both in the gain and in the offset and at both
wavelengths simultaneously, we should get a conservative estimate of
the uncertainties. In Fig.~\ref{fig:T_beta_plot}, the resulting
shifts of the observed ($T, \beta$) pairs are shown for the three
random pixels. The full maps of the maximum and minimum $\beta$ values
are shown in the Appendix, Sect.~\ref{sect:beta_appendix}.

\subsection{Dust opacity}  \label{sect:emissivity}

We calculated extinction maps at a 1.5$\arcmin$ resolution using stars
from the Two Micron All Sky Survey (2MASS, Skrutskie et al.
\cite{Skrutskie2006}) and the NICER method (Lombardi \& Alves
\cite{Lombardi2001}). The values of $A_{\rm V}$ are obtained from NIR
colour excess assuming an extinction law with $R_{\rm V}$=3.1. The
zero level of the extinction was set with the Schlegel et al.
(\cite{Schlegel1998}) extinction map. The extinction estimates enable
us to estimate the sub-millimetre dust opacity relative to the
visual extinction. The extinction is converted to total ISM mass with
the Bohlin et al. (\cite{Bohlin78}) relation, $N({\rm HI + H_2}) \sim
1.87 \times 10^{21} \, A_{\rm V}$ that is multiplied by 1.4\,$m_{\rm
H}$ to account for the mass of helium. The conversion is done for
easier comparison with the literature but the conversion factor itself
is appropriate mainly for diffuse medium and may vary across the
fields. 

\begin{figure*}
\centering
\includegraphics[width=!]{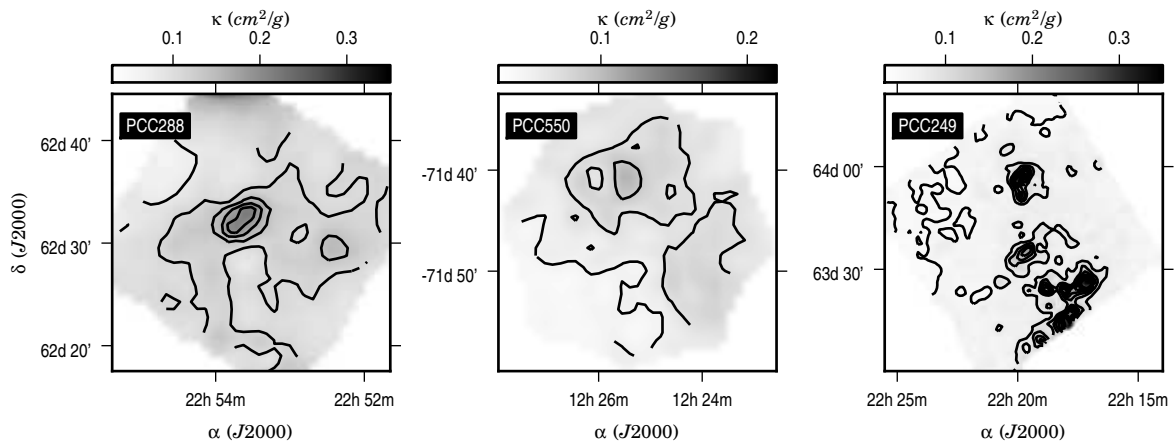}
\caption{
Maps of the dust opacity derived from the comparison of the NIR
extinction maps and SED fits with a fixed value of the spectral index,
$\beta=2.0$. The values are given for 250$\mu$m and relative to the
total ISM mass. The contours are drawn starting with the value of
0.05\,cm$^2$\,g$^{-1}$ and with a step of 0.02\,cm$^2$\,g$^{-1}$
(PCC550) or 0.03\,cm$^2$\,g$^{-1}$ (PCC288 and PCC249).
}
\label{fig:emissivity}
\end{figure*}

Figure~\ref{fig:emissivity} shows the 250\,$\mu$m opacity maps
derived from the SED fits done with $\beta=2.0$. The median
values in the fields PCC288, PCC550, and PCC249 are 0.11, 0.064, and
0.061\,cm$^2$\,g$^{-1}$. 
The opacities obtained with fixed value of $\beta=2.0$ and the free
$\beta$ are identical to within 20\%. In the central part of PCC288
the values rise close to $\kappa(250\,\mu{\rm m})\sim$0.22 while in
the northern colder region of PCC249 (corresponding to the Planck
detection P1) the peak value is even higher, above
$\sim0.3$\,cm$^2$\,g$^{-1}$. The value used for the mass estimates in
Sect.~\ref{sect:N} was 0.23\,cm$^2$\,g$^{-1}$. This is clearly above
the median values but roughly consistent with the values associated
with the highest column densities. However, in the field PCC550 the
values remain everywhere below $\kappa(250\,\mu{\rm
m})=$\,0.12\,cm$^2$\,g$^{-1}$. 
Our typical opacity of 0.1\,cm$^2$\,g$^{-1}$ corresponds to 
$\tau$(250$\mu$m)=2.3$\times 10^{-25}$/H, slightly more than twice the
value found in diffuse medium (Boulanger et al. \cite{Boulanger1996}).

\section{Radiative transfer model of PCC288} \label{sect:model}

To find out the level at which temperature variations along the
line-of-sight may bias the observed values of the spectral index and
the dust temperature, we made a simple model of the PCC288 region. 
The model was not intended as a fit to the PCC288 observations but is
only used to estimate the general observational biases in a setting
similar to this field. The model was based on the column density map
shown in Fig.~\ref{fig:colden} and represents a box with a size of
5.4\,pc (for $d$=800\,pc). The line-of-sight density profile was
gaussian with a FWHM that was 0.4\,pc towards the densest part of the
cloud and increased up to 1.6\,pc in regions of low column density.
The numbers roughly correspond to the scales of the central clump and
the whole field but are basically ad hoc values. In particular, the
density distribution of the model is quite smooth with a maximum
density of only 3.7$\times 10^4$\,cm$^{-3}$. Because also the
wavelength dependence of the spectral index could bias the observed
$\beta$ values (see Malinen et al.~\cite{Malinen2011}), the modelling
was done with a modified version of the normal Milky Way dust model
(Draine \cite{Draine2003}) where $\beta$ was set to a constant value
of 2.0 for all wavelengths $\lambda>100\mu$m. The cloud is illuminated
from the outside by an isotropic interstellar radiation field (ISRF;
Mathis \cite{Mathis83}). Three blackbody radiation sources with a
temperature of 2000\,K and luminosities of 10, 20, and 50\,$L_{\sun}$
were added to represent internal heating. We show the results for a
model where the column densities and the strength of the ISRF were
scaled so that the surface brightness levels match the observations of
PCC288 to $\sim 30\%$. 

\begin{figure}
\centering
\includegraphics[width=7.9cm]{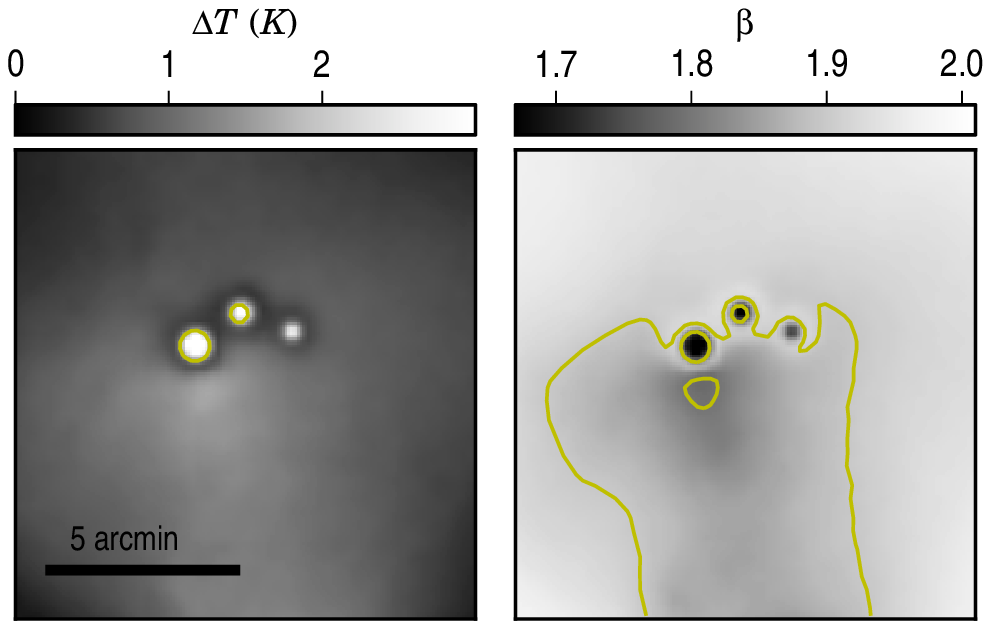}
\caption{
Results of the radiative transfer model of the PCC288 region. The left
hand frame shows the difference between the colour temperature derived
from synthetic observations and the real mass averaged dust
temperature in the model. The right hand frame shows the obtained
spectral index $\beta$. The contours are drawn at $\Delta T = 1\,$K
and $\Delta T = 2\,$K and at $\beta$=1.8 and $\beta=$1.9.
}
\label{fig:model}
\end{figure}

Because the proper handling of regions near the internal sources
requires high spatial resolution, the model was sampled onto a
hierarchical grid with the smallest cell size corresponding to the
resolution of 1024$^3$ uniform grid. The dust temperature
distributions were solved with a radiative transfer program that is
based on our Monte Carlo code (Juvela \& Padoan~\cite{Juvela2003}) but
has been adapted for use with hierarchical grids (Lunttila et al., in
preparation). The obtained surface brightness maps at 160\,$\mu$m,
250\,$\mu$m, 350\,$\mu$m, and 500\,$\mu$m were analysed to derive
colour temperature and spectral index maps. Noise was not added to the
synthetic observations and all wavelengths were given equal weight in
the ($T$, $\beta$) fitting. The results are shown in
Fig.~\ref{fig:model} where the three internal sources are clearly
visible (in order of increasing luminosity from West to East). The left
hand frame shows the difference between the observed colour
temperature and the mass weighted average of the true temperature. In
the region shown, the dust temperature is overestimated by between
$\sim$0.25\,K in the most diffuse areas and $\sim$8\,K at the position
of the strongest source. The median error is $\sim$0.8\,K. An error in
the temperature is associated with an opposite change in the spectral
index. In a fit with a fixed value of $\beta=2$, the median error
would be only 0.4\,K. At a temperature of $T=15$\,K, this would
correspond to a less than 15\% underestimation of column density. If
$\beta$ is kept as a free parameter, the 0.8\,K bias leads to a
$\sim$30\% error in column density when the estimate is based on
$\kappa$ at $160\,\mu {\rm m}$. This decreases to only $\sim$10\% when
the estimate is based on 500\,$\mu {\rm m}$. However, close to the
sources the errors remain large, at least a factor of two.

In Fig.~\ref{fig:model}, the median value of the observed spectral
index is $\beta$=1.92 which is only slightly smaller than the actual
value of the dust model, $\beta$=2.0. The values are more biased in
areas of high column density and the observed $\beta$ is consistently
below 1.9 in the south. At the position of the sources $\beta$ is
reduced further to a minimum value of 1.56. The observations provide
good estimates of the spectral index only in the more diffuse regions
but also at a certain distance from the sources where the internal
heating apparently helps to keep the temperature constant along the
line-of-sight. Although the column density is highest in the central
clump, the bias is higher in the south, further away from the heating
sources.

\section{Discussion}  \label{sect:discussion}

\subsection{Colour temperatures}

The calculation of the colour temperature maps for the full Herschel
fields confirmed the Planck detections of cold dust emission. Even in
the field PCC249, with active star formation and a high average colour
temperature, the Planck detections correspond to regions where the
colour temperatures, $T\sim15$\,K, are $\sim$2 degrees below the
values of the surrounding areas. The Planck detections are based on
the cold dust signature that remains when the warm component traced by
the 100\,$\mu$m IRIS maps has been subtracted (Montier et al.
\cite{Montier2010}). This does not require a low colour temperature
for the total emission. The separation of the warm and the possible
cold component is more uncertain in warm regions where the warm
component is strong and must be subtracted with higher relative
accuracy. The PCC249 sources are very bright with peak surface
brightness reaching in both PACS and the 250\,$\mu$m SPIRE band
several thousand MJy\,sr$^{-1}$. The Planck observations
were able to locate the two nearby cold regions despite the fact that,
according to Herschel, they are small and thus beam diluted: the
Planck beams have a significant contribution also from the warmer
surrounding regions.

In the analysis performed with a fixed spectral index value of
$\beta=2.0$, the minimum colour temperatures were 13.8\,K, 13.1\,K,
15.6\,K, and 14.8\,K for PCC288, PCC550, and the northern and the
southern areas in PCC249, respectively. These are a couple of degrees
higher than the values reported in Juvela et al. (\cite{Juvela2010})
where temperatures were calculated after the subtraction of the local
background. In spite of the different nature of the fields, the
minimum temperatures are rather similar. Of course, even the
background subtracted observations do not give a true picture of the
coldest temperatures that inside the cores are likely to be below
10\,K (e.g., Crapsi et al. \cite{Crapsi2007}; Harju et al.
\cite{Harju2008}). With radiative transfer modelling of the dust
continuum observations one could derive an upper limit for the central
temperature but once the temperature is reduced close to 6\,K, the
cold dust will be difficult to detect even with Herschel observations.
In our simple model of the field PCC288, the minimum physical dust
temperature was only $\sim$2\,K lower than the minimum colour
temperature. It is very likely that in the real clumps the central
densities rise higher than in our radiative transfer model and
therefore also reach lower physical temperatures.

\subsection{Cloud masses and dust opacity}

In Juvela et al. (\cite{Juvela2010}) the masses were estimated only
for compact objects, with a fixed circular aperture and with
background subtraction. The masses were quoted also for a 9 arcmin
diameter aperture and these can now be compared with the values of 
Table~\ref{table:masses} that were obtained without background
subtraction. In PCC288, the estimate has increased by a factor of two from
140\,$M_{\sun}$ to 288\,$M_{\sun}$. In the case of PCC550 the
background level is small and the change is much smaller. The previous
values of 9.2\,$M_{\sun}$ and 8.1\,$M_{\sun}$ for P1 and P2,
respectively, have been replaced by 12.2\,$M_{\sun}$ and
11.3\,$M_{\sun}$. In PCC249 the intensity levels are again high and
also the contrast between the emission in the selected apertures and
the surrounding reference areas is small. As a result, the masses
obtained without background subtraction are much larger. The previous
estimates of 90\,$M_{\sun}$ (area $P1$) and 93\,$M_{\sun}$ (area $P2$)
have increased to 252\,$M_{\sun}$ and 259\,$M_{\sun}$.  Differences
are, of course, not surprising because the new measurements are for
the whole mass along the line-of-sight while the old ones correspond
to the column density excess within the aperture.

In PCC249, the new values are in better agreement with mass estimates
from molecular line observations. Wang et al. (\cite{Wang2009})
obtained from $^{13}$CO(2 --1) observations a value of
1700\,$M_{\sun}$ for a region that corresponds to our area $P1$. The
estimate was calculated for a cloud distance of 1.67\,kpc which, with
the distance of $d=800$\,pc assumed here, translates to
$\sim$390\,$M_{\sun}$.  The value derived from Herschel data is still
lower but already $\sim$65\% of this number.
On the other hand, the situation has not changed much for PCC550 where
the masses obtained from molecular line studies (Vilas-Boas et al.
\cite{VilasBoas1994}; see Juvela et al. \cite{Juvela2010}) are also
higher than our values. In PCC550-P2 the molecular line data gave a
mass of 16.5\,$M_{\sun}$ and the $\sim$50\% discrepancy is still not
very serious. However, for the core P1 (core Mu5 in Vilas-Boas et al.
\cite{VilasBoas1994}) the CO analysis lead to a value of
$\sim33$\,$M_{\sun}$ that is three times our estimate. The uncertainty
of $\kappa$ can explain most of this discrepancy. The values derived
in Sect.~\ref{sect:emissivity} were all below $\kappa(250\,\mu {\rm
m})=0.15$\,cm$^2$\,g$^{-1}$ with a median value of
0.073\,cm$^2$\,g$^{-1}$. With $\kappa(250\,\mu{\rm
m})=0.085$\,cm$^{2}$\,g$^{-1}$ the mass estimate of P1 would be in
perfect agreement with the value obtained from the line observations.
The radiative transfer model of PCC288 hinted that the column density
estimates may be biased. In the case of PCC550 the final answer to
this question may be provided by the detailed radiative transfer
modelling that is currently under way (Ysard et al., in preparation). 
One can also question the accuracy of the CO mass estimates. The
fractional abundances are not accurately known, the observations of
the cold regions may be affected by some degree of molecular depletion
(e.g., PCC550), and, conversely, the optical depths of the employed
$^{13}$CO lines can become high enough so that radiative transfer
effects can bias the results (e.g., self-absorption affecting the
observed $^{13}$CO/C$^{18}$O line ratios). Considering all the sources
of uncertainty, the mass estimates obtained from the lines and from
dust continuum are in good agreement.

The median values of the dust absorption cross sections
$\kappa(250\mu{\rm m})$ obtained from the comparison with NIR
extinction are 0.13, 0.073, and 0.055\,cm$^2$\,g$^{-1}$ for the fields
PCC288, PCC550, and PCC249, respectively. The peak values are higher
by more than a factor of two and, in PCC550 and PCC249, are associated
with regions of higher column density. This behaviour is expected and
could be explained by an increase in grain sizes. In PCC288, there is
some indication that the dust opacity has increased in the central
clump, close to the column density maximum. The increase towards the
North corner is very likely only an artifact. The results could change
noticeably even within the limits set by the accuracy of the zero
points of the surface brightness scale. Nevertheless, the results
suggest spatial variations at least by a factor of two.  Most of our
$\kappa(250\,mu{\rm m})$ values are in the range
0.05--0.2\,cm$^2$\,g$^{-1}$. Scaled with $\nu^2$, this corresponds to
$\kappa(850\,\mu{\rm m})$=0.004--0.017\,cm$^2$\,g$^{-1}$. 
The $A_{\rm V}$ values were calculated assuming an extinction law with
$R_{\rm V}$=3.1. This is appropriate for most areas but not
necessarily for the densest cores. By adopting $R_{\rm V}$=5.5, the
extinction values would decrease by $\sim$16\%. 
Taking into account a possible decrease in the ratio $N(H)/A_{\rm V}$
the calculated dust opacities could increase further by more than 30\%
(see Evans et al. \cite{Evans2009}, Draine \cite{Draine2003}, and the
online
tables\footnote{http://www.astro.princeton.edu/$~$draine/dust/dust.html}).
This would further strengthen the case for an increasing
sub-millimetre dust opacity in the cold and dense cores.

The opacities can be compared with values found in other studies.
Using COBE measurements, Boulanger et al. (\cite{Boulanger1996})
derived an effective dust cross section for high latitude diffuse
medium. At 850\,$\mu$m, the obtained values are $\kappa(850\,\mu{\rm
m})$=0.005\,cm$^2$\,g$^{-1}$ in agreement with the low end of our
measurements. In the dark cloud IC~5146, Kramer et al.
(\cite{Kramer2003}) found the values to vary from
$\kappa_{850}/\kappa_{V} \sim 1.3\times 10^{-5}$ in warm regions
(20-30\,K) to $\sim 5 \times 10^{-5}$ in colder regions with
$T\sim12$\,K. With the conversion used in Sect.~\ref{sect:emissivity},
these correspond to $\sim$0.003\,cm$^2$\,g$^{-1}$ and
0.011\,cm$^2$\,g$^{-1}$, in general agreement with our range of
values. However, there are also dense and cold clouds where similar
increase of dust opacity has not been observed. Nutter et al.
(\cite{Nutter2008}) reported observations of a filament in Taurus and
concluded that these can be modelled accurately without any spatial
variations of dust properties. Juvela et al. (\cite{Juvela2009})
estimated for a Corona Australis molecular cloud filament a $\kappa$
value that was almost consistent with values found in diffuse clouds.
This in spite of the fact that the region had more than 20$^{\rm m}$
of visual extinction and had a minimum colour temperature of
$\sim$11\,K although, on the other hand, the sub-millimetre
measurements may have suffered from some spatial filtering. The
results should be tested, possibly with the help of Herschel
observations, to confirm whether the dust behaviour is this different
in apparently similar clouds or whether the results could be explained
by the difficulty of reaching an accuracy of measurements that is
needed for a reliable determination of dust opacities.

\subsection{Spectral indices}  \label{sect:disc_beta}

We fitted the observed ($T$, $\beta$) values of Fig.~\ref{fig:T_beta}
with two analytical formulas, $\beta=A (T/20{\rm K)}^{-\alpha}$ and
$\beta=(\delta+\omega T)^{-1}$. The parameters of the relations are
listed in Table~\ref{table:alpha}. The obtained $T(\beta)$ relations
are steeper than reported in Dupac et al. (\cite{Dupac2003}) who
fitted PRONAOS observations using the latter formula. Our relation is
closer to the results from the Archeops experiment (D\'esert et al.
\cite{Desert2008}) where the observations fit a relation $\beta=A
T^{-\alpha}$ with an exponent value of $\alpha=0.66\pm0.054$. 
The error estimates in Table~\ref{table:alpha} do not reflect the true
uncertainty but only the accuracy of the curve fitting. One source of
uncertainty is the accuracy to which the very small grain contribution
could be subtracted from the 100\,$\mu$m data. However, the
temperature and spectral index maps derived without the 100\,$\mu$m
band are relatively similar to the results shown in
Fig.~\ref{fig:T_beta} (see Appendix~\ref{sect:T_beta_no_100}). We will
next examine the reliability of the detection of the spectral index
variations in the presence of other error sources.

Calibration errors affect the ($T$, $\beta$) values mostly in a
systematic fashion. In Fig.~\ref{fig:T_beta_plot}, we indicated for
three representative pixels the shifts that would result from a 10\%
change in the gain calibration combined with a change in the offset
that is equal to the uncertainties listed in
Table~\ref{table:offsets}.  The calibration errors tend to move all
points along a trajectory similar to that caused by random noise
without much affecting the detection of an inverse $T-\beta$ relation.
In the field PCC550, the spectral index reaches values $\beta
\sim$2.35 at the locations of the two cores. In the high-$\beta$
version (calibration errors causing higher 250\,$\mu$m and lower
500\,$\mu$m surface brightness) the centre of the PCC550 filament
would reach $\beta \sim 2.7$ while the background would remain at
$\beta \sim $2.4. Opposite changes in the calibration would decrease
the spectral index values in the filament to a value $\sim$2, still
0.2--0.3 units above the background. This shows that the increase of
dust opacity within the PCC550 filament is robust against calibration
errors although the uncertainty in the absolute value of $\beta$ is
high, close to $\sim$0.4. The effects of a different calibration are
similar in the other two fields, changing the absolute level of
$\beta$ without a clear change in the morphology.

We used Monte Carlo simulations to quantify the noise bias and the
evidence for a negative $\beta-T$ correlation. We added noise to the
surface brightness measurement and repeated the fit of the $\beta(T)=A
T^{-\alpha}$ relation.The statistical uncertainty of the surface
brightness measurements, $\sigma_{\rm S}$, was estimated from the
difference between the observed maps and the SED fits with a free
$\beta$ parameter. These are conservative estimates because the
residuals are affected also by calibration errors. The calibration
uncertainty $\sigma_{\rm C}$ was assumed to consist of a 10\%
uncertainty in the gain calibration and the zero point uncertainties
listed in Table~\ref{table:offsets}. To see how sensitive the
extracted values are on the statistical noise alone, we carried out
one set of simulations where $\sigma_{\rm S}$ was added to the
observed maps. However, we mainly compare the original observations
with reference simulations with a flat $\beta(T)$ relation. 
These are based on the SED fits that were made with $\beta=2.0$. In
the first set of simulations $\sigma_{\rm S}$ and $\sigma_{\rm C}$
were added to the input maps. In the second set of simulations one
quarter of each surface brightness map was additionally scaled with a
random factor corresponding to a 10\% relative error. This simulates
mapping artifacts, $\sigma_{\rm M}$, whose effect is fundamentally
different from both $\sigma_{\rm S}$ (affecting each pixel
independently) and $\sigma_{\rm C}$ (affecting all pixels
simultaneously). While $\sigma_{\rm C}$ can shift all the ($T$,
$\beta$) points along almost a fixed trajectory, without a noticeable
change in the best fitting $\beta(T)$ relation, the mapping artifacts
will increase the dispersion of the ($T$, $\beta$) values along the
noise-induced $\beta(T)$ curve. This could be misinterpreted as a
stronger evidence of an inverse relation between $\beta$ and $T$. The
comparison between the SED fits and the surface brightness maps
suggested the presence of some such artifacts (see
Appendix~\ref{sect:residual_maps}). In PCC249, these were below
$\sim$5\% apart from the northern end of the 160\,$\mu$m map where,
within an area less than 5\% of the whole map, the residual rises
between 10\% and 20\%. In PCC288 the artifacts remain below 10\% when
the northern corner has been masked out. The same applies to PCC550
except for the 160\,$\mu$m map where the difference to the SED fit is
above 10\% over about one third of the map and reaches values 20--30\%
over an area less than 10\% of the full map. By introducing a 10\%
error in the quarter of the mapped area and in all the frequency
bands, we are certain to obtain a conservative estimate of the mapping
artifacts, especially in PCC288 and PCC249.

\begin{figure}
\centering
\includegraphics[width=7.9cm]{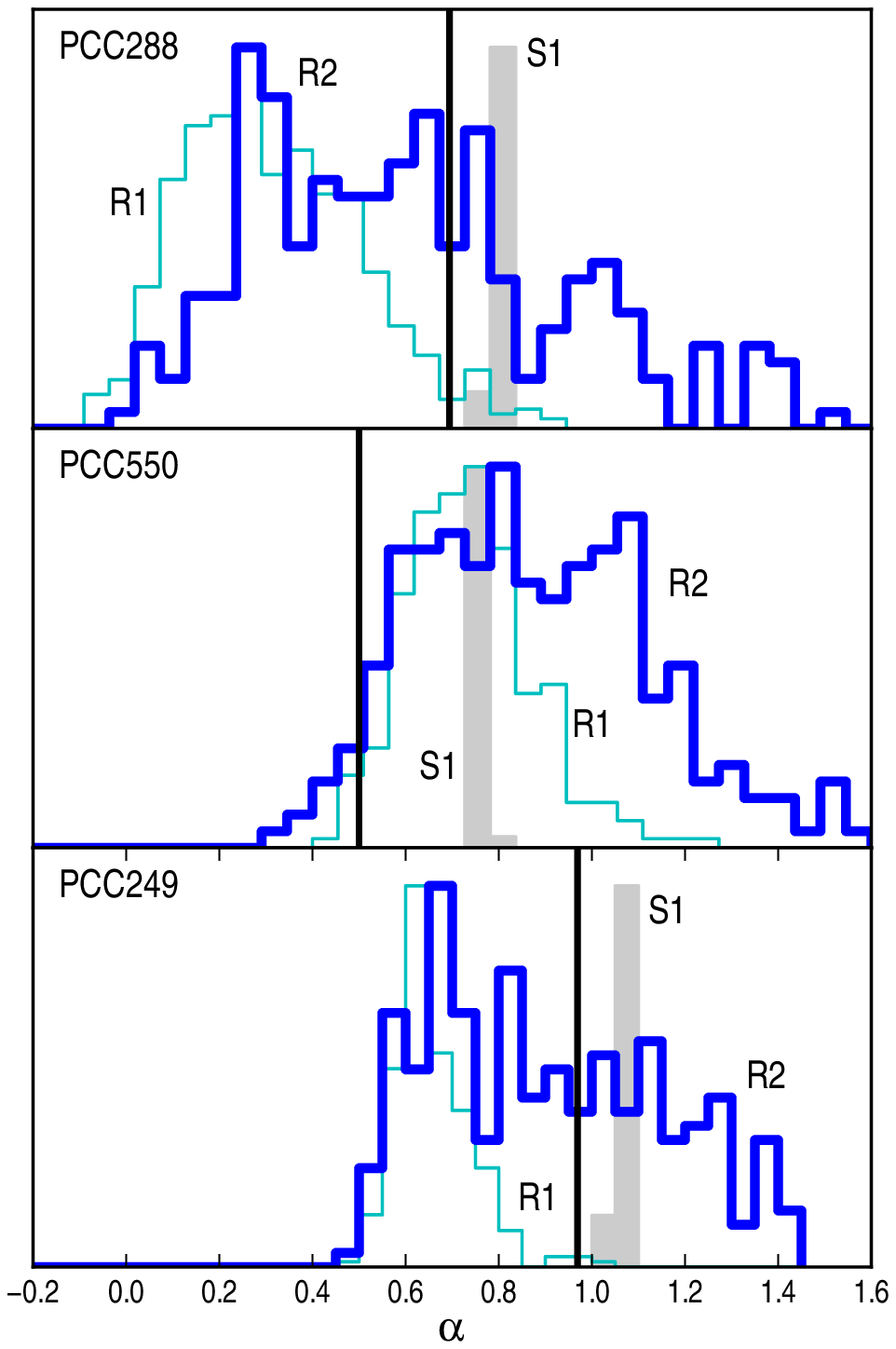}
\caption{
Results of the Monte Carlo study where the distribution of the values ($T$,
$\beta$) is modelled with a function $\beta(T)=A T^{-\alpha}$.
The vertical lines show the values of the parameter $\alpha$ estimated
from the observations. The histograms $S1$ shows the distribution of
the $\alpha$ values for observations with the added noise $\sigma_{\rm
S}$. These are compared to simulations with $\beta=$2.0 containing the
noise components $\sigma_{\rm S}$ and $\sigma_{\rm C}$ (distribution
$R1$) or with additional simulated mapping artifacts, $\sigma_{\rm M}$
($R2$, see the text for details).
}
\label{fig:MC}
\end{figure}

Figure~\ref{fig:MC} shows the distribution of the parameters $\alpha$
in this Monte Carlo study, each resulting from the simulation of 500
sets of synthetic surface brightness maps. For each set of frequency
maps, the colour temperatures and the spectral indices were estimated
pixel by pixel, and the analytical $\beta(T)$ formulas were fitted to
the results. The distribution $S1$ corresponds to the observed maps
with added noise, $\sigma_{\rm S}$. The distributions $R1$ and $R2$
correspond to simulations with a constant value of $\beta=2$, $R1$
containing $\sigma_{\rm S}$ and $\sigma_{\rm C}$ and $R2$ additionally
the mapping errors $\sigma_{\rm M}$. The vertical lines show the
parameter values obtained directly from the observations.

Because of the noise-induced anticorrelation between $T$ and $\beta$,
the $\alpha$ values are positive even when the input data correspond
to a constant value of $\beta=2.0$. For the same reason, the
distributions $S1$ are shifted relative to the observed values and are
narrow only because the effect is almost identical in all Monte Carlo
realizations.
In the fields PCC288 and PCC249, the observed $\alpha$ values are
higher than most values in the $R1$ distribution. If mapping artifacts
can be ignored, this indicates a real anticorrelation between
temperature and the observed spectral index. The hypothesis $\alpha>0$
gets probabilities 97\% and 99\% in the case of PCC288 and PCC249,
respectively. However, with the simulated mapping artifacts the
reference distributions $R2$ is much wider and the corresponding
probabilities are reduced to 69\% and 66\%. The detections are no
longer statistically significant but the numbers depend critically on
the actual level of $\sigma_{\rm M}$. We believe that the noise values
used in the simulation were conservative and the truth lies somewhere
between the cases $R1$ and $R2$. Therefore, the Monte Carlo study
gives some, although weak evidence for a non-flat $\beta(T)$ relation.
In the case of PCC550, the values of $\alpha$ were higher in the
simulations than in observations. This is possible if the mapping
artifacts conspired to produce a low $\alpha$ value or may indicate
overestimation of the errors.

\begin{table}
\caption{Parameters of the $\beta(T)$ fits.}
\label{table:alpha}  
\centering         
\begin{tabular}{l c c c c}
\hline\hline            
            &   \multicolumn{2}{c}{$\beta(T)=A (T/20{\rm K})^{-\alpha}$}  &
               \multicolumn{2}{c}{$\beta(T)=(\delta + \omega T)^{-1}$}
               \\
Target      &    $A$  & $\alpha$  & $\delta$ & $\omega$  \\
\hline         
 PCC\,288  &   1.89(0.02) &  0.62(0.03)  &   0.18(0.02)  &  0.018(0.001) \\
 PCC\,550  &   1.90(0.02) &  0.44(0.02)  &   0.26(0.02)  &  0.014(0.001) \\
 PCC\,249  &   1.77(0.01) &  0.97(0.02)  &   0.02(0.01)  &  0.027(0.001) \\
\hline
\end{tabular}
\end{table}

The Monte Carlo study showed that the global analysis of ($T$,
$\beta$) points does not give a conclusive proof of spectral index
variations. That analysis is not sensitive to very localized $\beta$
variations while it may be significantly affected by artifacts that
are more likely to exist close to map edges and far from our dense
sources.
In Fig.~\ref{fig:T_beta}, many of the spectral index variations are
clearly correlated with actual cloud structures and, therefore, are
unlikely to be caused by mapping artifacts. In PCC288 the spectral
index increases from $\sim$2.2 at the boundaries to $\sim$2.5 in the
central clump before decreasing to $\sim$1.5 at the location of the
strongest warm source at the centre of the map. In PCC249 the warm
sources are similarly associated with spectral index minima,
$\beta\sim$1.0 near PCC249-P1 and $\sim$1.3 near PCC249-P2, while the
values in the central map otherwise are between 1.8 and 2.2. It is
clear that warm sources are associated with significantly lower
$\beta$. The evidence for high $\beta$ values ($>>$2) towards the
coldest clumps is somewhat weaker because it comes from larger scales
that might be affected by mapping artifacts. However, in PCC288 and
PCC550, $T$ and $\beta$ are clearly anticorrelated also in the regions
with $T<$15\,K. In PCC249 the situation is similar at large scales 
but not in the cores PCC249-P1 and PCC249-P2. These are very near to
strong radiation sources which may have affected the spectral index
estimates.
 
We conclude that there is strong evidence of spatial variations in the
apparent dust spectral index and the magnitude of these changes is
likely to be at least $\Delta \sim 0.2$ and can be up to $\Delta \sim
1$ towards warm, internally heated sources. The simple model of the
PCC288 field showed that these apparent variations do not necessarily
indicate an equal change in the dust properties. In the model, the
spectral index of the grains could be 0.2--0.3 units higher than what
is observed (Sect.~\ref{sect:model}). This applies to regions of high
column density and without internal heating. The effect is opposite to
the observed correlation between the column density and $\beta$ and
further supports the idea of higher $\beta$ values for the dust grains
in the cold cores. This behaviour is expected on the basis of
laboratory measurements and can be explained by the properties of the
grain materials at low temperature (Mennella et al.
\cite{Mennella1998}; Boudet et al. \cite{Boudet2005}; M\'eny et al.
\cite{Meny2007}). On the other hand, the low values of $\beta$ that
were observed towards the warm sources in PCC288 can be fully
explained by line-of-sight temperature variations. It will be
difficult to determine whether also the intrinsic spectral index of
the grains is lower in such regions (see also Malinen et al.,
\cite{Malinen2011}). 
 
The exact relation between the observed and real $\beta$ depends on
the properties of the object in question (density structure, strength
of the external radiation field, nature of the internal sources etc.)
and the results of our radiative transfer model cannot be generalized
very far. We have started detailed modelling of the observations of
PCC550. The results of that study will give us a better idea on how
the observed spectral index variations, if real, are related to the
actual dust properties. This may also shed new light on the
reliability of the mass estimates derived from Herschel observations
(Ysard et al., in preparation).

The possible presence of mapping artifacts was found to be the main
problem in the evaluation of the observational $T$--$\beta$ relation.
In new observations larger map sizes and increased redundancy would
help to reduce such artifacts and make it easier to evaluate their
impact. The situation may be improved with better data reduction
methods and, in the future, one could also use Planck data to check
for low resolution artifacts in the long wavelength Herschel data.
When combined with far-infrared data, the Planck observations
themselves are very suitable for the study of the spectral index
variations and will be used to examine the situation also in cold
cores and clumps (Planck Collaboration \cite{Planck2011a},
\cite{Planck2011b}).

\section{Conclusions}  \label{sect:conclusions}

We have studied Herschel surface brightness measurements of three
areas where the Planck satellite data indicated the presence of
significant amounts of cold dust. By analysing surface brightness maps
at wavelengths between 100\,$\mu$m and 500\,$\mu$m we map the dust
colour temperature, dust opacity, and dust spectral index
variations over the fields. The results lead to the following
conclusions:
\begin{itemize}
\item The colour temperature maps confirm the Planck
detections of cold dust. The analysis of the total intensity leads to
minimum colour temperature of 13--15\,K in the three fields.
\item The results show an increase in the dust opacity
towards colder and denser cloud regions. The range of values for the
dust absorption cross section was estimated to be $\kappa(250\,\mu{\rm
m}) \sim$ 0.05--0.2\,cm$^2$\,g$^{-1}$, in agreement with previous
studies.
\item  The median values of the observed spectral 
indices were between 1.9 and 2.2. We witnessed an inverse relation
between the colour temperature and the spectral index. A Monte Carlo
study indicated but did not conclusively prove a temperature
dependence of $\beta$ beyond what could be produced by observational
errors. However, because the changes are well correlated with column
density, we believe the variations to be real. Very low values down to
$\beta\sim$1.0 were observed towards warm sources but these are likely
to be caused mainly by temperature variations rather than changes in
the grain characteristics.
\item With the help of a simple radiative transfer model, we examined the 
relation between the observed and intrinsic spectral indices. The
results show that the intrinsic $\beta$ value of the dust grains is
underestimated in regions of high column density and, therefore, the
observations may also underestimate the real variation of the spectral
index. On the other hand, the flat spectrum near internal heating
sources can be explained by the line-of-sight temperature variations
without any change in the dust properties.
\end{itemize}

\appendix

\section{Frequency maps} \label{sect:frequency_maps}

\begin{figure}
\centering
\includegraphics[width=8cm]{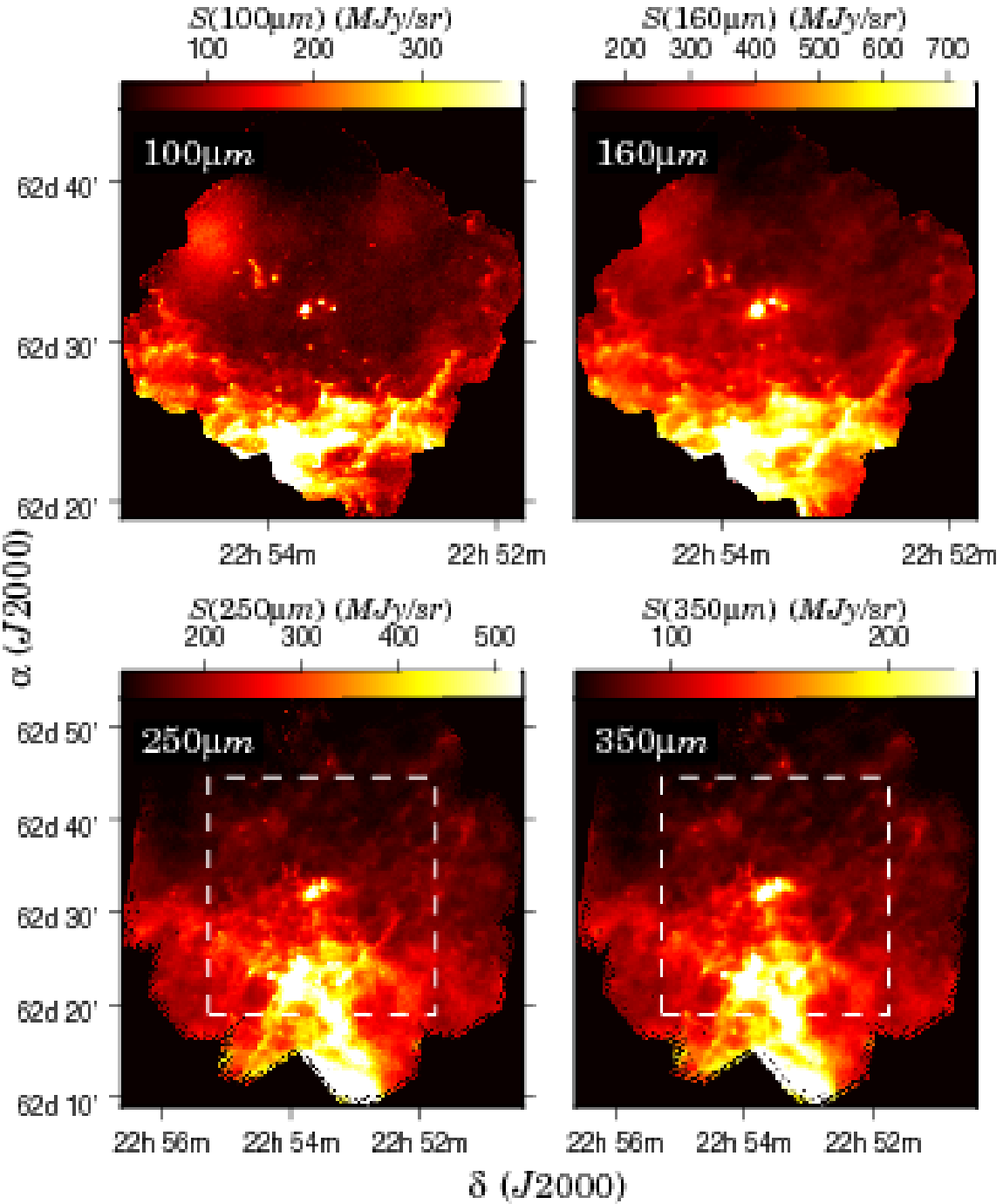}
\caption{
Surface brightness maps of the field PCC288. The zero point of the
intensity scale has been set by comparison with the IRIS and Planck
maps. In the lower frames the dashed line corresponds to the size of
the upper frames showing the PACS observations.
}
\label{fig:maps_PCC288}%
\end{figure}

\begin{figure}
\centering
\includegraphics[width=8cm]{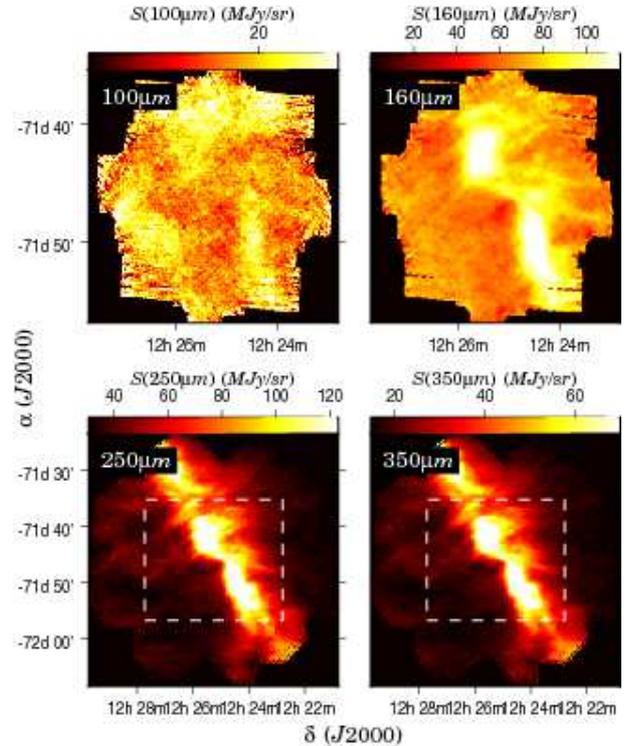}
\caption{
Surface brightness maps of the field PCC550.
}
\label{fig:maps_PCC550}%
\end{figure}

\begin{figure}
\centering
\includegraphics[width=8cm]{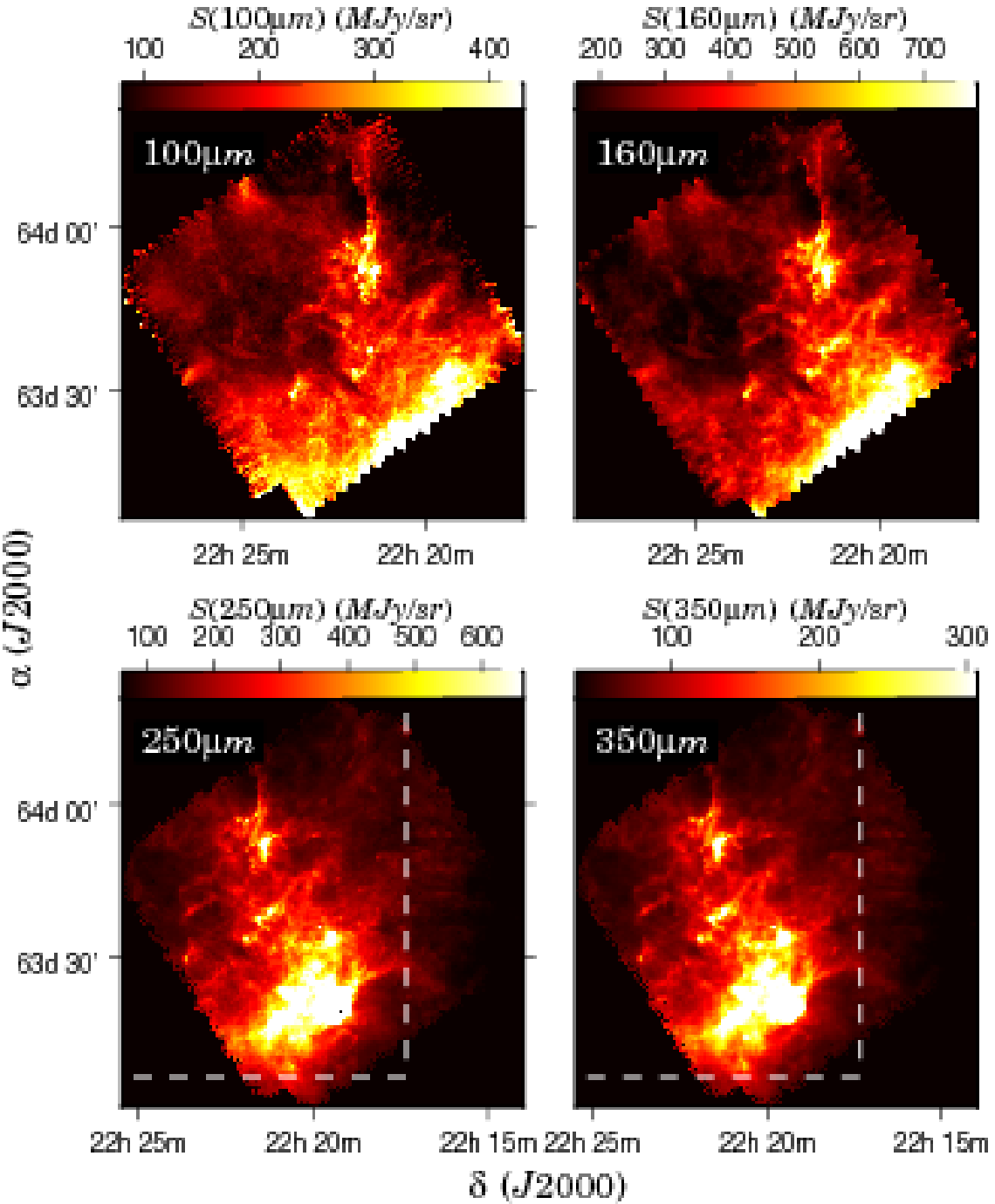}
\caption{
Surface brightness maps of the field PCC249.
}
\label{fig:maps_PCC249}%
\end{figure}

Figures~\ref{fig:maps_PCC288}--\ref{fig:maps_PCC249} show the
100\,$\mu$m, 160\,$\mu$m, 250\,$\mu$m, and 350\,$\mu$m surface
brightness maps for the three SDP fields. The maps for 500\,$\mu$m are
not included but their appearance, apart from the lower intensity and
lower spatial resolution, is almost identical with the 350\,$\mu$m
maps.

\section{Residual maps at 160\,$\mu$m--500\,$\mu$m} \label{sect:residual_maps}

\begin{figure}
\centering
\includegraphics[width=8cm]{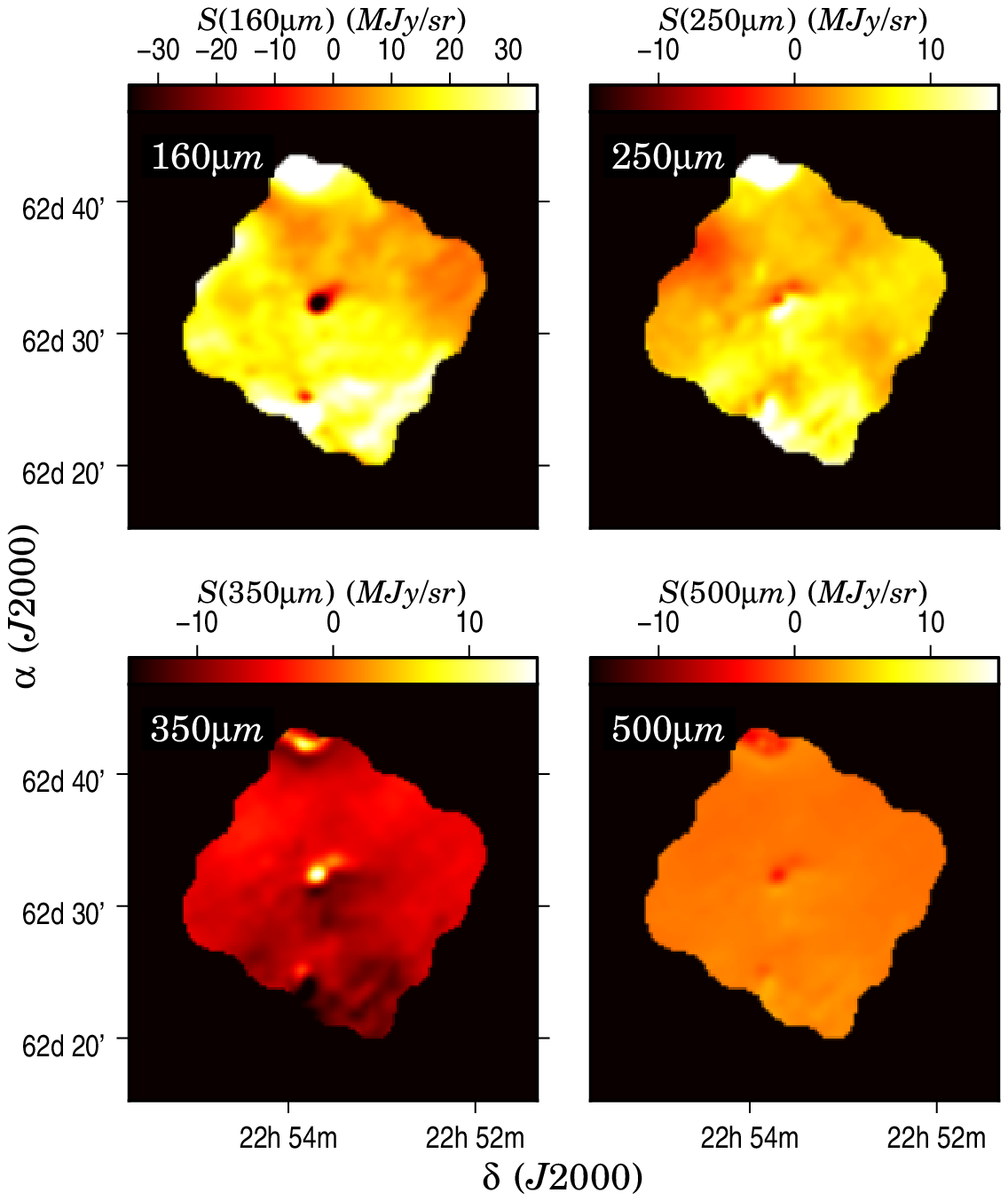}
\caption{
Residual maps of the surface brightness in the field PCC288 at four
wavelengths when observations were fitted with the spectral index as a
free parameter. The resolution is 1 arc minute.
}
\label{fig:maps_PCC288_residuals}%
\end{figure}

\begin{figure}
\centering
\includegraphics[width=8cm]{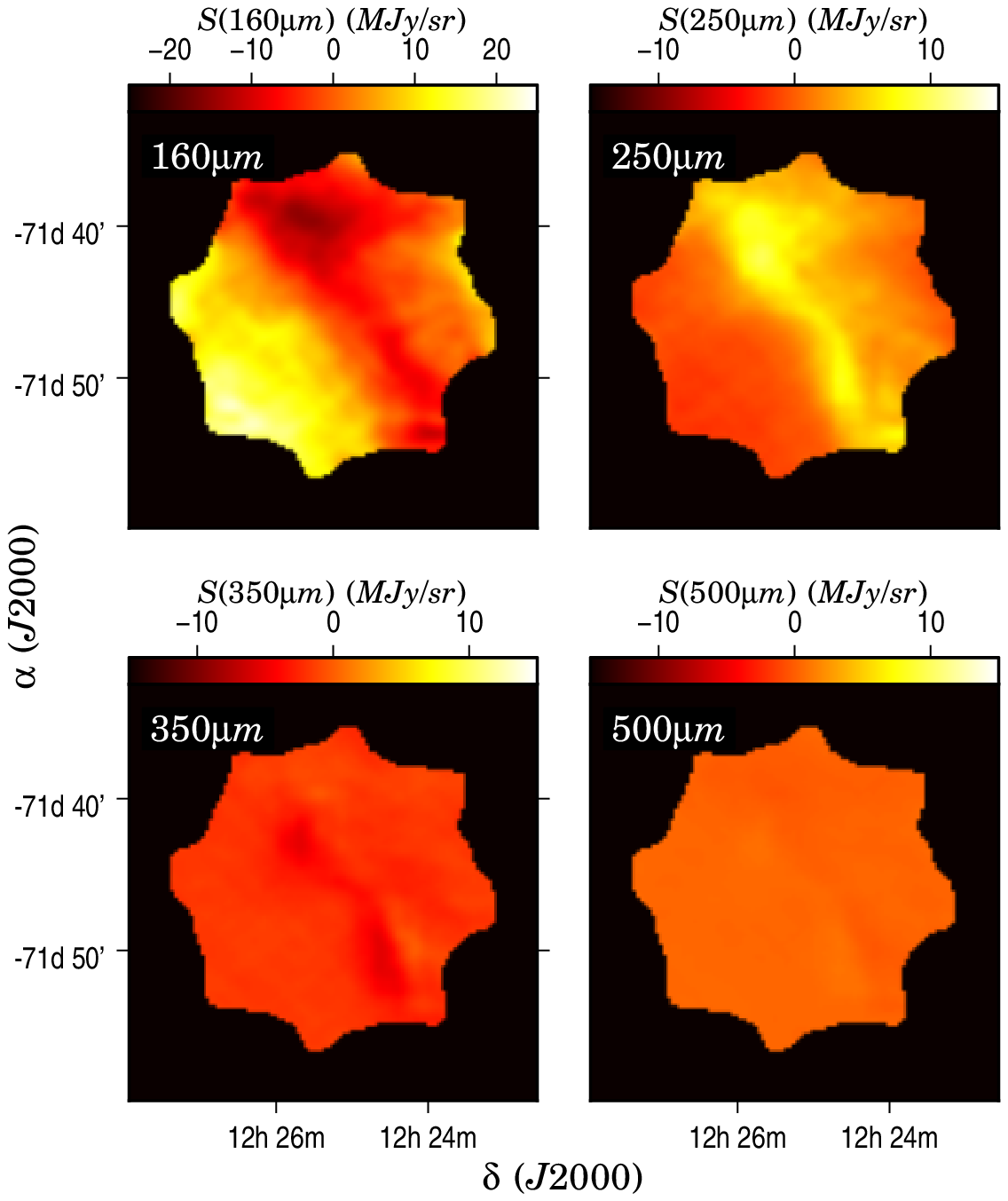}
\caption{
Residual maps of the surface brightness in the field PCC550 at 
four wavelengths.
}
\label{fig:maps_PCC550_residuals}%
\end{figure}

\begin{figure}
\centering
\includegraphics[width=8cm]{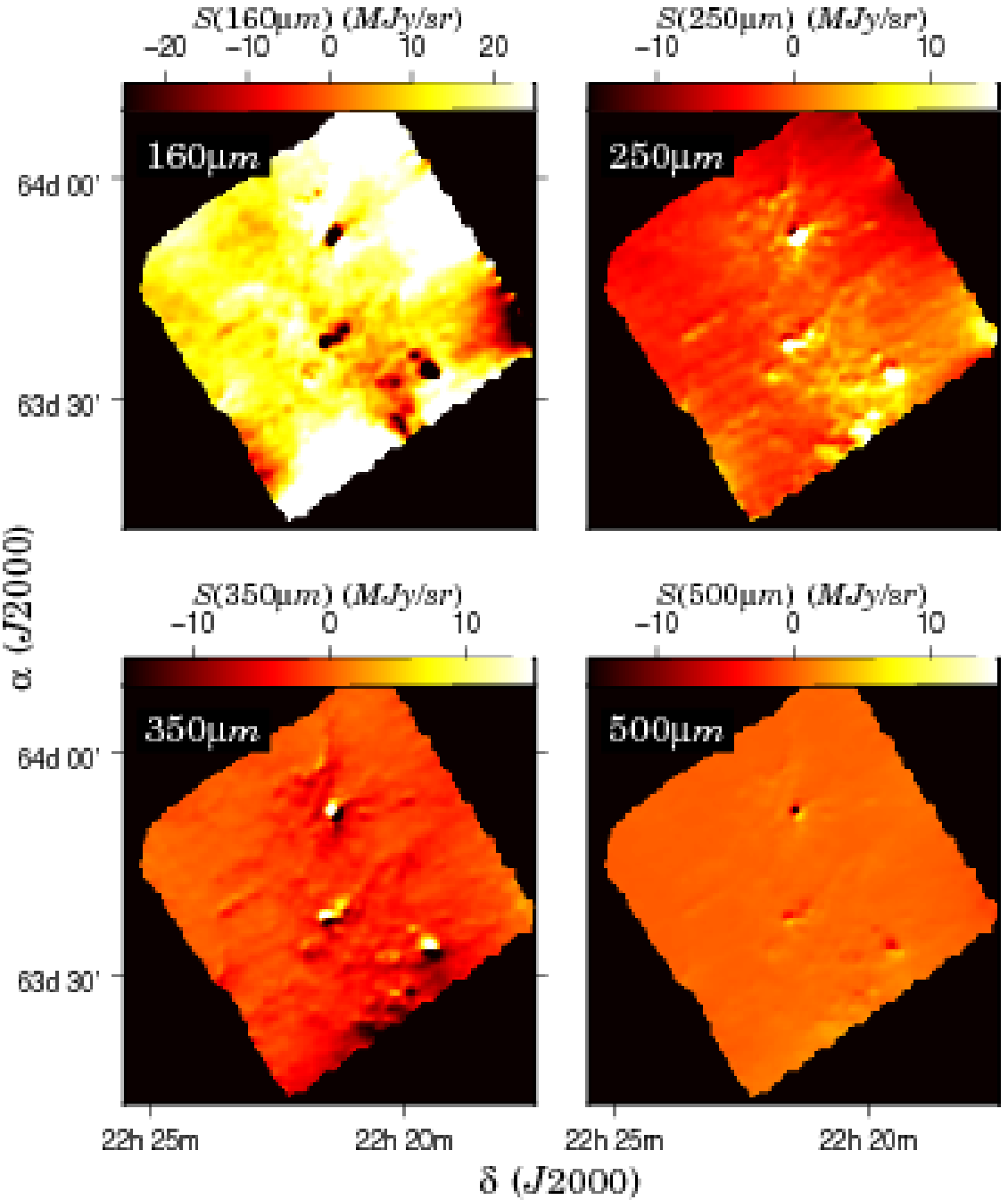}
\caption{
Residual maps of the surface brightness in the field PCC249 at the
four wavelengths.
}
\label{fig:maps_PCC249_residuals}%
\end{figure}

The quality of the $B_{\nu}(T) \nu^{\beta}$ fits and the maps
themselves can be evaluated by examining the residual maps at the five
wavelengths used in the SED fit. These maps are shown in
Figs.~\ref{fig:maps_PCC288_residuals}--\ref{fig:maps_PCC249_residuals}.

In PCC288 the average residuals are $-4.0$, 16.9, 6.7, $-6.3$, and
1.4\,MJy\,sr$^{-1}$ in the five bands in the order of increasing
wavelength. The corresponding values are 
$-$0.35, 2.4, 2.4, $-1.9$, and 0.4\,MJy\,sr$^{-1}$ in the field PCC550
and 
$-3.8$, 17.3, $-0.7$, $-1.6$, and 0.47\,MJy\,sr$^{-1}$ in the field
PCC249.
Allowing some errors in the gain, the estimated accuracies of the
offsets (see Table~\ref{table:offsets}) appear to have been realistic.

In PCC288 the largest residuals, $\sim 100$\,MJy\,sr$^{-1}$, are found
at the centre of the 160\,$\mu$m map. This is associated with the
brightest point source and is still less than 10\% of the surface
brightness. In Fig.~\ref{fig:maps_PCC249_residuals} the residual
images are strongly saturated and one can see both negative and
positive values on the two sides of a source. Although the residuals
reach several hundred MJy\,sr$^{-1}$ these are only at a $\sim$5\%
level. The positions of the sources themselves agree in all frequency
maps so that astrometric errors should not be a significant factor.
Unlike in the case of the fast scanning speed, the intermediate
scanning speed used in our observations should not result in a
significant smearing of the PACS beam. Nevertheless, the residuals
could be associated with an imperfect deconvolution of the detector
time constants.

\subsection{Maps of the spectral index uncertainty}
\label{sect:beta_appendix}

\begin{figure}
\centering
\includegraphics[width=8cm]{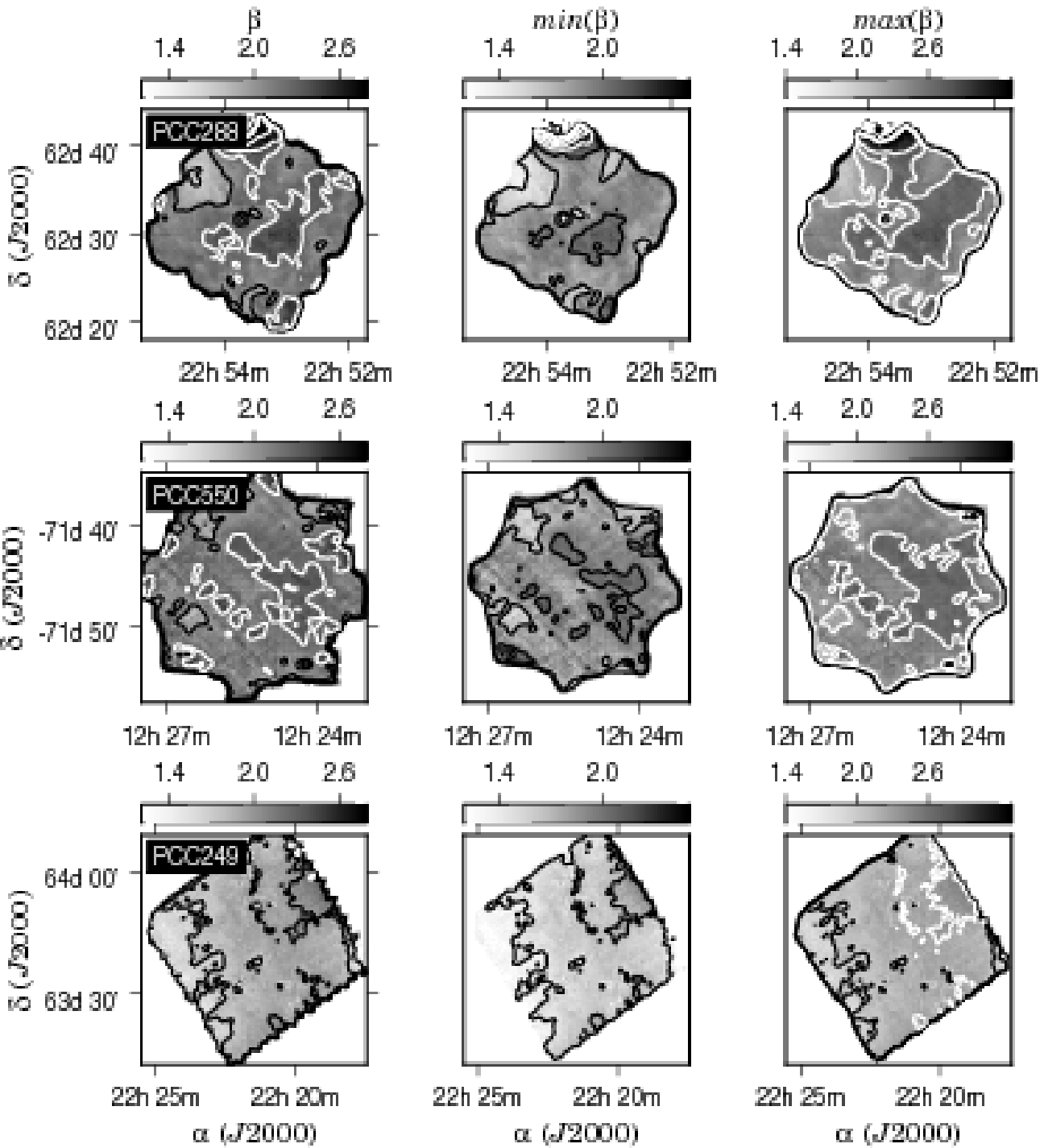}
\caption{
Estimated uncertainty of the spectral index maps. The first column
shows the derived values (as in Fig.~\ref{fig:T_beta}).  The second
and the third columns show the minimum and maximum values of $\beta$
consistent the estimated zero point uncertainties and an error of 10\%
in the gain calibration. The contours are drawn between 1.4 and 2.9 in
steps of 0.3 units (white contours starting with value 2.3).
}
\label{fig:beta_error_maps}%
\end{figure}

The derived spectral index values are affected by observational noise,
errors in the zero point of the adopted surface brightness scale, and
errors in the gain calibration. 
Figure~\ref{fig:beta_error_maps} shows maps that can be used to assess
the importance of gain calibration and zero point errors. As discussed
in Sect.~\ref{sect:beta}, we modified the 250\,$\mu$m and 500\,$\mu$m
surface brightness maps by adding or subtracting the zero point
uncertainties from Table~\ref{table:offsets} and by scaling the maps
by an amount corresponding to a 10\% gain uncertainty. The changes at
the two wavelengths were made in opposite directions in order to have
the maximum effect in the $\beta$ value. By applying 1$\sigma$ shifts
in the gain and in the offset and at both wavelengths critical for the
determination of the spectral index we should get conservative limits
for the spectral indices. The maps of minimum and maximum spectral
indices are shown in Fig.~\ref{fig:beta_error_maps}.

\subsection{Temperature and spectral index maps without 100\,$\mu$m
data} \label{sect:T_beta_no_100}

\begin{figure}
\centering
\includegraphics[width=8cm]{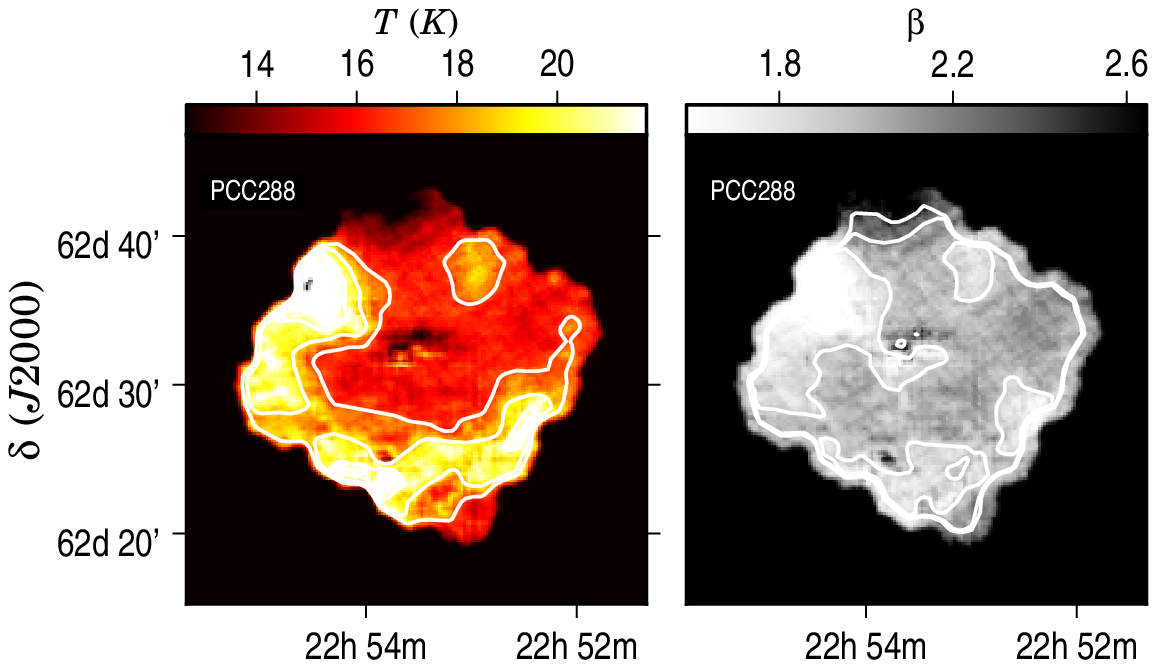}
\includegraphics[width=8cm]{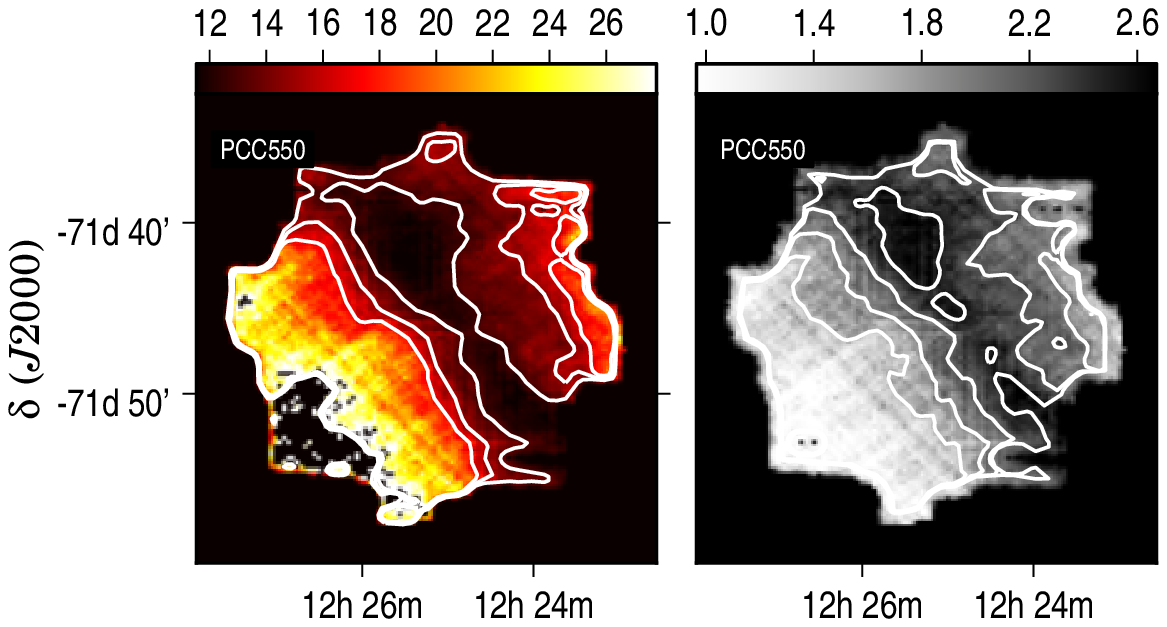}
\includegraphics[width=8cm]{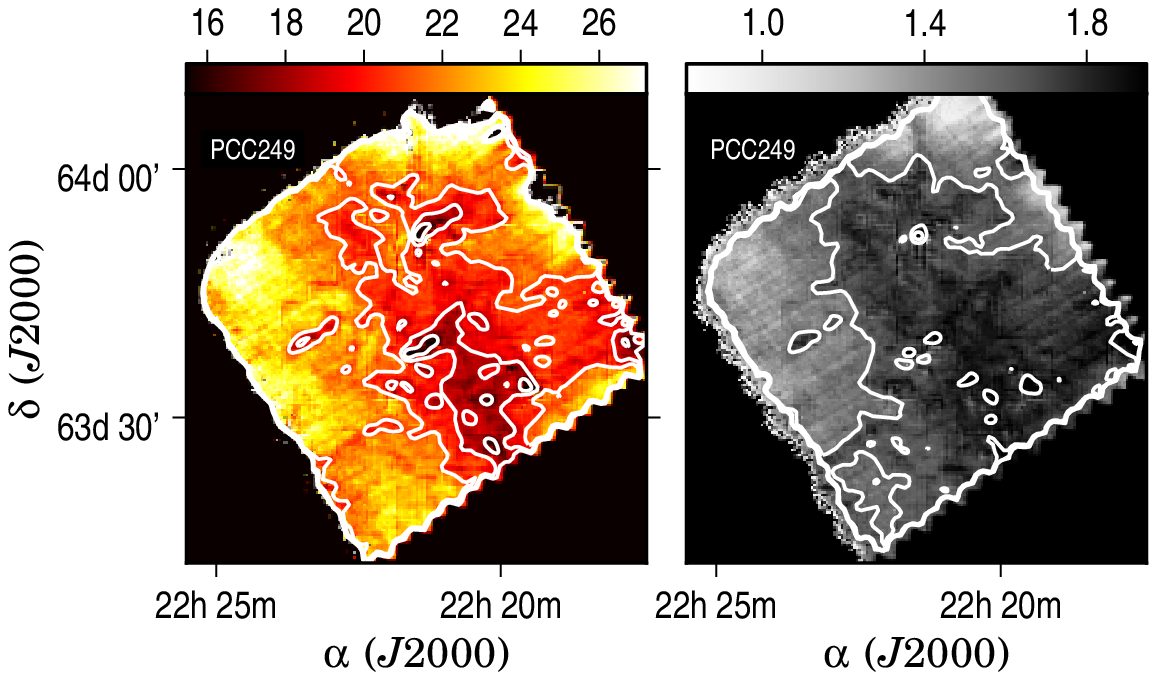}
\caption{
The colour temperature and dust spectral index map obtained by 
fitting the observations between 160\,$\mu$m and 500\,$\mu$m.
}
\label{fig:T_beta_no_100}
\end{figure}

The subtraction of the VSG emission from the 100\,$\mu$m data relied
on the ratio of the VSG and BG grains in the D\'esert et al.
(\cite{Desert1990}) dust model. This also affects the 160\,$\mu$m data
but only as an uncertainty of the zero point of the intensity scale.
To check to what extent the colour temperature and the spectral index
maps are influenced by the 100\,$\mu$m data, we recomputed the
parameters using the 160\,$\mu$m--500\,$\mu$m data only. The resulting
maps of the colour temperature and the spectral index are shown in
Fig.~\ref{fig:T_beta_no_100}.

The maps are morphologically similar to those shown in
Fig.~\ref{fig:T_beta}. The largest differences are in PCC550 where,
without the 100\,$\mu$m data, the spectral index map closely follows
the shape of the dense filament. The appearance of SW part of that map
again suggests the presence of some artifact, the $\beta$ values
decreasing close to one. Compared to Fig.~\ref{fig:T_beta}, PCC249
exhibits somewhat lower values of $\beta$ and a higher range of
temperatures.

\begin{acknowledgements}
MJ and JM acknowledge the support of the Academy of Finland Grant No.
127015. JM acknowledges the support from V\"ais\"al\"a foundation.
This publication makes use of data products from the Two Micron All
Sky Survey, which is a joint project of the University of
Massachusetts and the Infrared Processing and Analysis
Center/California Institute of Technology, funded by the National
Aeronautics and Space Administration and the National Science
Foundation.
\end{acknowledgements}

\end{document}